\documentclass[11pt]{article}
\usepackage{mathrsfs}
\usepackage{natbib,graphicx,setspace,lscape,longtable}
\usepackage{mathrsfs,amsmath,amsthm,amssymb,color}
\usepackage{natbib,epsfig,graphicx,pdfpages}
\usepackage{rotating}
\usepackage{subfigure}
\usepackage{graphicx}

\bibpunct{(}{)}{;}{a}{,}{,}

\newtheorem{theorem}{Theorem}
\newtheorem{lemma}{Lemma}
\newtheorem{proposition}{Proposition}
\newtheorem{remark}{Remark}
\def\beq{\begin{equation}}
\def\eeq{\end{equation}}
\def\beqr{\begin{eqnarray}}
\def\eeqr{\end{eqnarray}}
\def\beqrs{\begin{eqnarray*}}
\def\eeqrs{\end{eqnarray*}}
\def\bet{\begin{theorem}}
\def\eet{\end{theorem}}
\def\bel{\begin{lemma}}
\def\eel{\end{lemma}}
\def\bep{\begin{proposition}}
\def\eep{\end{proposition}}
\def\bg{\begin{figure}[tbph]\begin{center}}
\def\eg{\end{center}\end{figure}}

\def\bc{\begin{center}}
\def\ec{\end{center}}

\def\wt{\widetilde}
\def\wh{\widehat}

\def\my{\mathbf y}

\def\mR{\mathbb{R}}

\def\vu{\mathbf u}
\def\bJ{\mathbf J}

\def\var{\mbox{var}}
\def\cov{\mbox{cov}}

\def\RSS{\mbox{RSS}}

\numberwithin{equation}{section}

\newcommand{\Var}{\textnormal{Var}}
\newcommand{\Cov}{\textnormal{Cov}}

\newcommand{\bA}{{\mathbf A}}
\newcommand{\bB}{{\mathbf B}}
\newcommand{\bF}{{\mathbf F}}

\newcommand{\bG}{{\mathbf G}}
\newcommand{\bH}{{\mathbf H}}
\newcommand{\bI}{{\mathbf I}}
\newcommand{\bK}{{\mathbf K}}

\newcommand{\bQ}{{\mathbf Q}}
\newcommand{\bP}{{\mathbf P}}
\newcommand{\bR}{{\mathbf R}}
\newcommand{\bS}{{\mathbf S}}
\newcommand{\bU}{{\mathbf U}}
\newcommand{\bV}{{\mathbf V}}
\newcommand{\bW}{{\mathbf W}}

\newcommand{\ba}{{\mathbf a}}
\newcommand{\bb}{{\mathbf b}}

\newcommand{\be}{{\mathbf e}}

\newcommand{\bu}{{\mathbf u}}
\newcommand{\bv}{{\mathbf v}}
\newcommand{\bw}{{\mathbf w}}
\newcommand{\bx}{{\mathbf x}}
\newcommand{\by}{{\mathbf y}}
\newcommand{\bz}{{\mathbf z}}

\newcommand{\bbeta}  {\boldsymbol{\beta}}

\newcommand{\bSigma}{\boldsymbol{\Sigma}}

\newcommand{\bve}{\mbox{\boldmath$\varepsilon$}}

\newcommand{\btheta} {\boldsymbol{\theta}}
\newcommand{\bxi} {\boldsymbol{\xi}}

\newcommand{\bD}{{\mathbf D}}

\newcommand{\ve}{{\varepsilon}}
\renewcommand{\epsilon}{{\ve}}
\renewcommand{\hat}{\widehat}
\def\wt{\widetilde}

\begin{document}

\title{\bf Banded Spatio-Temporal Autoregressions}

\author{
Zhaoxing Gao$^1$ \quad
Yingying Ma$^2$ \quad Hansheng Wang$^3$ \quad Qiwei Yao$^{1,3}$\\
$^1$Department of Statistics, London School of Economics\\
$^2$School of Economics and Management, Beihang University\\
$^3$Guanghua School of Management, Peking University
}

% \date{}

\maketitle

\begin{abstract}
We propose a new class of spatio-temporal models with unknown and banded autoregressive
coefficient matrices. The setting represents a sparse structure for high-dimensional
spatial panel dynamic models when panel members represent
economic (or other type) individuals at many different locations.
The structure is practically meaningful when the order of panel members is arranged
appropriately. Note that the implied autocovariance matrices are unlikely to be banded,
and therefore, the proposal is radically different from the existing
literature on the inference for high-dimensional banded covariance matrices.
Due to the innate endogeneity, we apply the least squares method based on a Yule-Walker
equation to estimate autoregressive coefficient matrices. The estimators based on multiple Yule-Walker equations are also studied.  A ratio-based
method for determining
the bandwidth of autoregressive matrices is also proposed. Some asymptotic properties
of the inference methods are established.
The proposed methodology is further illustrated
using both simulated and real data sets.
\end{abstract}


\newpage

\section{Introduction}

One common feature in most literature on spatial econometrics is to
specify each autoregressive coefficient matrix in a spatial autoregressive or
a spatial dynamic panel model as a product of an unknown scalar parameter and
a known spatial weight matrix, and the
 focus of the inference is on those a few unknown
scalar parameters placed in front of spatial weight matrices. See,
for example, Cliff and Ord (1973), Yu et al. (2008), Lee and Yu (2010),
Lin and Lee (2010), Kelejian and Prucha (2010), Su (2012), and Yu et al. (2012).
Using spatial weight matrices reflects the initial thinking that
spatial dependence measures should take into account both spatial
locations and feature variables at locations simultaneously.
A weight matrix may reflect the closeness of different spatial locations. It needs
to be specified subjectively.  There are multiple weighting
possibilities including inverse distance, fixed distance, space-time
window, $K$-nearest neighbors, contiguity, and spatial interaction.
The conceptualization specified in spatial matrices for
a particular analysis imposes a specific structure onto the data collected
across the locations. Ideally one would select a conceptualization
that best reflects how the features actually interact with
each other in the real world.

For a given application it is not always obvious how to specify a
pertinent spatial weight matrix. Consequently
the resulting spatial autoregressive model may be incapable to accommodate adequately
the dependent structure across different locations. Dou et al. (2016) considers the models
% of Yu et al. (2008, 2012)
which employ different scalar coefficients, in front of spatial weight
matrices, for different locations. By drawing energy and inspiration
from the recent development in sparse high-dimensional (auto)regressions
(Guo et al. 2016), we propose in this paper a new class of
spatio-temporal models in which autoregressive coefficient matrices are
completely unknown  but are assumed to be banded, i.e. the non-zero coefficients only occur within the narrow band around the
main diagonals. This avoids the difficulties in specifying spatial weight  matrices
subjectively.
The setting specifies autoregressions over neighbouring locations only.
The underpinning idea rests on the fact that in many
applications it is enough to collect information from neighbouring
locations, and then the information from farther locations become redundant.
Of course the banded structure relies on arranging all the locations concerned in a
unilateral order. In practice, an appropriate  ordering can be
deduced from subject
knowledge aided by statistical tools such as cross-validation; see Section 4.2.
It is worth pointing out that the implied autocovariance matrices are
unlikely to be banded in spite of the banded autoregressive coefficient matrices.

Guo et al. (2016) considered banded autoregressive models for vector time series,
and estimated the coefficient matrices by a componentwise least squares method.
Unfortunately their method does not apply to our setting, due to the endogeneity
in spatial autoregressive models. Instead we adapt a version of generalized method of moments estimation based on
a Yule-Walker equation (Dou et al. 2016). % Unlike Dou et al. (2016),
Furthermore the estimation of the parameters based on multiple
Yule-Walker equations is also investigated.
The asymptotic property of the estimation is established when the
dimensionality $p$ (i.e. the number of panels) diverges together with the
sample size $n$ (i.e. the length of the observed time series). The convergence rates of the estimators are the same with those in Dou et al. (2016).
More precisely, the estimated coefficients are asymptotically normal when
$p = o(\sqrt{n})$, and is consistent when $p=o(n)$.

In practice, the width of the nonzero coefficient bands in the coefficient matrices
needs to be estimated. We propose a ratio-based estimation method which is shown to
lead to a consistent estimated  width when both $n$ and $p$ tend to infinity.

The rest of the paper is organized as follows. We specify the class of models and
the associate estimation methods in Section 2. The asymptotic properties are presented
in Section 3. The numerical illustration with both simulated and real data sets are
reported in Section 4. All technical proofs are relegated into an Appendix.

\section{Model and estimation method}
\subsection{Spatio-temporal regression model}
 Consider the spatio-temporal regression
\begin{equation}
\label{eqn:BSAR}
\my_t = \bA \my_t + \bB \my_{t-1} + \bve_t,
\end{equation}
where $\my_t = (y_{1,t},...,y_{p,t})^{\top}$ represents the
observations collected from $p$ locations at time $t$,
$\bve_t=(\epsilon_{1,t},\epsilon_{2,t},..., \epsilon_{p,t})^{\top}$
is the innovation at time $t$ and  satisfies the condition that
\[E(\bve_t) =0, \qquad \Var(\bve_t) =\bSigma_\ve \quad \text{and}\qquad \cov(\my_{t-j},\bve_t)=0 \;\; {\rm
for \; all} \; j\ge 1,\]
where $\bSigma_\ve$ is an unknown positive definite matrix.
Furthermore we assume that $\bA \equiv  (a_{i,j})$ and
$\bB \equiv  (b_{i,j})$ are $p\times p$ unknown banded coefficient matrices, i.e.,
\begin{equation} \label{b0}
a_{i,j} = b_{i,j} = 0 \ \  \textnormal{for all} \ \ |i-j| > k_0,
\end{equation}
and $a_{i,i}=0$ for $1\le i \le p$.
We call $k_0$ ($<p$) the bandwidth parameter which is an unknown positive integer.
In the above model (\ref{eqn:BSAR}),  $\bA$  captures the pure  spatial
dependency among different locations, and $\bB$ captures the dynamic dependency.

Model (\ref{eqn:BSAR}) extends the popular spatial dynamic panel data models (SDPD) substantially.
The standard SDPD assumes that each coefficient matrix is a product of a known linkage matrix
and an unknown scalar parameter,  {see, e.g.,  \cite{yu2008quasi} and \cite{yu2012}. } While some sparse structure has to be
imposed in order to conduct meaningful inference when $p$ is large, the
inflexibility of having merely single parameter in each regression
coefficient matrix is too restrictive, see, e.g., \cite{dou2016generalized}.
Note that the condition $a_{i,j}=b_{i,j}=0$ %for all $|i-j|>k_0$
does not imply Cov$(y_{i,t},
y_{j,t})=0$ or Cov$(y_{i,t}, y_{j,t-1}) =0$, regardless of the covariance
structure of $\bve_t$, see (\ref{b1}) below. Instead the banded sparse structure
imposed in (\ref{eqn:BSAR}) implies that  conditionally on the
information among the `closest neighbours', the information from farther
locations become redundant. This reflects the common sense in many
practical situations, though the definition of the closeness is case-dependent.

Let $\bI_p-\bA$ be invertible,
and all the eigenvalues of $(\bI_p - \bA)^{-1} \bB$ be smaller than $1$ in modulus,
where $\bI_p$ denotes  the $p \times p $ identity matrix.
 Then model (\ref{eqn:BSAR}) can be rewritten as
\begin{equation} \label{b2}
 \my_t = (\bI_p - \bA)^{-1} \bB\my_{t-1} + (\bI_p - \bA)^{-1} \bve_t,
\end{equation}
which admits a (weakly) stationary solution of  $ \my_t$.
For this stationary process, $E\by_t =0$, and the Yule-Walker equations are
\begin{equation} \label{b1}
\bSigma_0 = (\bI_p - \bA)^{-1} \bB \bSigma_1^\top + (\bI_p - \bA)^{-1}\bSigma_\ve (\bI_p - \bA^\top)^{-1},
\quad \bSigma_j = (\bI_p - \bA)^{-1} \bB \bSigma_{j-1} \;\; {\rm for} \;\; j\ge 1,
\end{equation}
where $\bSigma_j= \cov(\my_{t+j}, \my_t)$ for any $j\geq 0$.
Since the inverse of a banded matrix is unlikely to be banded, $\bSigma_0$, therefore also
$\bSigma_j$ are not banded in general. We refer to \S 4.3 of \cite{golub2013matrix}, and  \cite{kilicc2013inverse} for the properties and the computation of banded matrices and their inverses.

Throughout this paper, $\my_t$ is referred to as a stationary process defined by (\ref{b2}).

\subsection{ Generalized Yule-Walker estimation}
As $\my_t$ appears on both sides of equation \eqref{eqn:BSAR} and $\my_t$ is
correlated with $\bve_t$, the least squares estimation based on
regressing $\my_t$ on $(\my_t, \my_{t-1})$
directly leads to inconsistent estimators, due to the innate endogeneity of \eqref{eqn:BSAR}.
We observe that the second equation of (\ref{b1}) implies
\begin{equation} \label{YWE}
 \bSigma_1^\top \be_i =  \bSigma_1^\top \ba_i + \bSigma_0 \bb_i \equiv \bV_i \bbeta_i,\quad i=1,...,p,
\end{equation}
where $\be_i$ denotes the $p\times 1$ unit vector with 1 as its $i$-th element,
$\bA^\top = (\ba_1, \cdots, \ba_p)$, $ \bB^\top = (\bb_1, \cdots, \bb_p)$,
$\bbeta_i$ is the $\tau_i \times 1$ vector
obtained by stacking together the non-zero elements in $\ba_i$ and $\bb_i$, and
$\bV_i$ is the $p \times \tau_i$ matrix consisting of the corresponding columns of
$ \bSigma_1^\top$ and $\bSigma_0$.
It follows from  (\ref{b0}) that
\begin{equation}
\tau_i \equiv \tau_i(k_0) = \left\{
\begin{array}{ll}
2(k_0+i)-1 & 1 \le i \le k_0,\\
4k_0 + 1 & k_0 < i \le p - k_0, \\
2(k_0 +p-i) +1 \qquad & p-k_0 < i \le p.
\end{array}
\right.
\end{equation}

We first treat the bandwidth $k_0$ as a known parameter and apply a version of generalized method of moment estimation based on (\ref{YWE}), i.e.
% In the spirit of Dou et al. (2016),
we apply least squares method to estimate $(\bA,\bB)$ by solving the
following minimization problems
\begin{equation}
\mathop{\min}_{\ba_i,\bb_i}\|\wh \bSigma_1^\top \be_i -\wh \bSigma_1^\top \ba_i - \wh \bSigma_0 \bb_i\|_2^2,\quad i=1,...,p,
\end{equation}
where
\begin{equation} \label{b4}
\wh \bSigma_1 = \frac{1}{n} \sum_{t=2}^{n} \my_t \my_{t-1}^\top \ \quad  \text{and}\quad \ \
\wh \bSigma_0 = \frac{1}{n} \sum_{t=2}^{n} \my_{t-1} \my_{t-1}^\top.
\end{equation}
We omit the term $\my_n \my_n^T$ in the definition of $\wh \bSigma_0$ above for a minor
technical convenience which ensures the validity of (2.11) and  (\ref{b5}) below. Let $\hat{\bz}_i=\wh \bSigma_1^\top \be_i $ and $\wh \bV_i$ be the sample version of $\bV_i$ in (2.5), (2.7) leads to
 the least square estimator
\begin{equation}\label{eqn:YWest}
\wh \bbeta_i = (\wh \bV_i^\top\wh \bV_i)^{-1} \wh \bV_i^\top \wh \bz_i, \quad i =1,..., p.
\end{equation}
The corresponding residual sum of squares is
\begin{equation} \label{b7}
\RSS_i \equiv \RSS_i(k_0) = {1 \over p} \| \wh \bz_i - \wh \bV_i \wh \bbeta_i\|^2, \quad i =1,..., p.
\end{equation}
We note that (2.10) is a function of $k_0$, while in practice, $k_0$ is unknown and we will propose a consistent way to estimate $k_0$ in Section 2.4 below.

Combining all the estimators in (\ref{eqn:YWest}) together leads to the
estimators for $\bA$ and $\bB$,  which are denoted by,
respectively, $\wh \bA$ and $\wh \bB$.

It follows from (\ref{eqn:BSAR}), (\ref{b0})  and (\ref{b4}) that
\begin{equation} \label{b13}
\wh \bz_i = {1 \over n} \sum_{t=2}^n \by_{t-1} y_{i,t} = {1 \over n} \sum_{t=2}^n
\by_{t-1} ( \by_t ^\top \ba_i + \by_{t-1}^\top \bb_i + \epsilon_{i,t})
=\wh \bV_i \bbeta_i +  {1 \over n} \sum_{t=2}^n \by_{t-1} \epsilon_{i,t}.
\end{equation}
Hence it holds that
\begin{equation} \label{b5}
\wh \bbeta_i - \bbeta_i = {1 \over n} (\wh \bV_i^\top\wh \bV_i)^{-1} \wh \bV_i^\top \sum_{t=2}^n \by_{t-1} \epsilon_{i,t},
\quad i =1,..., p.
\end{equation}
In the above expressions,
\begin{equation} \label{b6}
\wh \bV_i = {1 \over n} \sum_{t=2}^n \by_{t-1} \bu_{t,i}^\top,
\end{equation}
where $\bu_{t,i}$ is a $\tau_i \times 1$ vectors consisting of $y_{j,t}$ for $j \in S_i$ and
$y_{\ell,t-1} $ for $\ell \in S_i^+$, where
\[
S_i=\{j: 1 \leq j \leq p, \ 1 \leq |j-i| \leq k_0  \} \quad\text{and}\quad
 S^+_i=\{j: 1 \leq j \leq p, \  |j-i| \leq k_0  \}.
\]

\subsection{ A root-$n$ consistent estimator for large $p$}
By Theorem 2 in Section 3 below,
the estimator (\ref{eqn:YWest}) admits a convergence rate
different from $\sqrt{n}$ when $p/\sqrt{n}\rightarrow \infty$.
This is an over-determined case in the sense that the number of estimation
equations is far greater than the number of parameters to be estimated.
 Similar results can also be found in  Dou et
al. (2016) and \cite{chang2015high}, among others. Borrowing the idea from Dou et al.
(2016), we propose an alternative estimator, which reduces the number of
the estimation equations from $p$ to a smaller constant. The resulting estimator
restores the $\sqrt{n}$-consistency  and is also asymptotically normal.
% The details of this new estimation method is  introduced as follows.

Note that, the $\ell$-th row of $\wh \bV_i$ is $\be_{\ell}^\top\wh \bV_i$. By
(\ref{b6}), this can be further expressed as ${1 \over n}\be_{\ell}^\top
\sum_{t=2}^n \by_{t-1} \bu_{t,i}^\top,$
which is the sample covariance between $ y_{l,t-1}$ and $\bu_{t,i}.$
Then, the strength of the correlation between  $y_{\ell,t-1}$ and  $\bu_{t,i} $  can be measured by
\begin{equation} \label{b10}
\delta_{\ell}^{(i)} = {1 \over n} \sum_{t=2}^n \Big(\sum_{j\in  S_i}|y_{\ell,t-1} y_{j,t}|+\sum_{j\in  S^+_i} |y_{\ell,t-1} y_{j,t-1}|\Big ).
\end{equation}
When $\delta_{\ell}^{(i)}$ is close to 0, the $\ell$-th equation in
(2.5) carries little information on $\bbeta_i$. Since our
concern is the estimation for $\bbeta_i$, we may only keep the $\ell$-th
equation  in (2.5) and hence (2.7) with the $d_i$ largest $\delta_{\ell}^{(i)}$.

Let $\bw_{t-1}^i \in d_i \times 1$ be the sub-vector of $\by_{t-1}$. Specifically, $\bw_{t-1}^i$ consists of those $y_{\ell,t-1}$ with the $d_i$ largest $\delta_{\ell}^{(i)}$. Then, we can obtain the new estimator as
\begin{equation} \label{b11}
\wt \bbeta_i = ( \wt\bW_i^\top\wt\bW_i)^{-1}  \wt\bW_i^\top \wt \bz_i, \quad i =1,..., p,
\end{equation}
  where
 \begin{equation} \label{b12}
   \wt\bW_i={1 \over n} \sum_{t=2}^n \bw_{t-1}^i \bu_{t,i}^\top \quad\mbox{and}\quad \wt \bz_i={1 \over n} \sum_{t=2}^n \bw_{t-1}^i y_{i,t}.
\end{equation}
Therefore,
\begin{equation*}
\wt \bbeta_i-  \bbeta_i={1 \over n} (\wt \bW_i^\top \wt \bW_i)^{-1}\wt \bW_i^\top \sum_{t=2}^n \bw_{t-1}^i \epsilon_{i,t},
\quad i =1,..., p.
\end{equation*}
Theorem 3 in Section 3 shows the asymptotic normality of the above estimator provided that the number of estimation equations used satisfies condition $d_i=o_p(\sqrt{n})$. In practice, $d_i$ should be a prescribed number and Theorem 3 is valid as long as the condition $d_i=o_p(\sqrt{n})$ holds uniformly for all $i$.

\subsection{Determination of bandwidth parameter $k_0$}

In practice, the bandwidth parameter $k_0$ is unknown. We propose below a method to
estimate it. Similar ideas can be found in \cite{lam2011estimation} and \cite{lam2012factor} for
 determining the number of factors in time series factor modelling.
%\noindent
%{\bf A generalized BIC criterion}: For $i=1, \cdots, p$ and  $k=1, \cdots, K$, let
%\begin{equation} \label{b8}
%\BIC_i(k) = \log \{ \RSS_i(k) +c_0 \} + \tau_i(k){ \log(p \vee n) \over n},
%\end{equation}
%where $ \tau_i(k)$ and $ \RSS(k)$ are defined, respectively, in (\ref{b3})
%and (\ref{b7}),  and $c_0 >$ is a small constant. We define an estimator for $k_0$ as
%\begin{equation} \label{b9}
%\wh k = \max_{1 \le i \le p} \big\{ \arg\min_{1 \le k \le K} \BIC_i(k) \big\}.
%\end{equation}

%\noindent
%{\bf Remark 1}. We insert a small positive constant $c_0$ in the
%logarithmic function in the definition of $\BIC_i(k)$ above such that this term
%will not converge to negative infinity.
%Since $\RSS_i(k)$ is the residual sum of squares based on `regression model'
%(\ref{eqn:YWith}) which does not contain a random error term.  When $k\ge k_0$, it is a valid
%sample version of Yule-Walker equation. Hence $\RSS_i(k) \to 0$  in
%probability as $n\to \infty$, for any $k \ge k_0$. Therefore
%\[
%\log \{ \RSS_i(k_0)\} - \log\{ \RSS_i(k_0+1)\} \approx
%\{ \RSS_i(k_0) - \RSS_i(k_0+1)\} \big/ \RSS_i(k_0+1)
%\]
%may be very large even when the decrease in residual, i.e. $\RSS_i(k_0) - \RSS_i(k_0+1)$,
%is neglibible. In practice we may set $c_0$ equal to 0.2 times the
%sample variance of $z_i = \wh \bSigma_1^T \be_i$. See also BICC of An et al (2008) for
%linear regression with large $p$ small $n$.
Let $K \ge 1$ be a known upper bound of $k_0$. Our estimation method is based on
the following simple observation:
 If we replace $(\wh \bSigma_0, \wh \bSigma_1)$ in (2.7)
by the true $( \bSigma_0, \bSigma_1)$, the corresponding
true value of $\RSS_i(k) $ is positive and finite for $1 \le k < k_0 $,
and is equal to 0 for $k_0 \le k \le K $. Thus the ratio $\RSS_i(k-1)\big/ \RSS_i(k)$
is finite for $k< k_0$, $\RSS_i(k_0-1)\big/ \RSS_i(k_0)$ is excessively large, and
$\RSS_i(k-1)\big/ \RSS_i(k) $ is effectively `0/0' for $k> k_0$.

To avoid the singularities when $k>k_0$, we introduce a small factor $w_n
={C}/{n}$ in the ratio for some constant $C>0$.
% and this is just for the theoretical purpose.
A ratio-based estimator for $k_0$ is defined as
 \begin{equation}
\label{eqn:ratio}
\wh k = \max_{1\le i \le p}\arg \max_{1< k \le
K} \frac{\RSS_i(k-1) + w_n }{\RSS_i(k) +  w_n},
\end{equation}
where $K\geq 1$ is a prescribed integer. Our numerical study shows that the procedure is insensitive to the choice of $K$ provided that $K\geq k_0$. In practice, we often choose $K$ to be $[n^{1/2}]$ or choose $K$ by checking the curvature of the ratio in (2.17) directly.

\subsection{Estimation with multiple Yule-Walker equations}
{In Section 2.3, we have established a $\sqrt{n}$-consistent estimator for $\bbeta_i$ with fewer estimation equations. However, this does not necessarily improve the estimation accuracy since we only make use of partial information for the parameters.} In Dou et al. (2016), the estimation of the parameters is based on only one Yule-Walker equation. In view of the equations in (\ref{b1}), we may also estimate $(\bA,\bB)$ using more than one Yule-Walker equations, {and therefore we have more information for $\bA$ and $\bB$}.  Let $r$ be a prescribed positive integer, we consider the following $r$ Yule-Walker equations:
\begin{equation}\label{myk}
\left(\begin{array}{c}
\bSigma_1^\top\\
\bSigma_2^\top\\
\vdots \\
\bSigma_r^\top
\end{array}\right)=\left(\begin{array}{c}
\bSigma_1^\top\\
\bSigma_2^\top\\
\vdots \\
\bSigma_r^\top
\end{array}\right)\bA^\top+\left(\begin{array}{c}
\bSigma_0^\top\\
\bSigma_1^\top\\
\vdots \\
\bSigma_{r-1}^\top
\end{array}\right)\bB^\top.
\end{equation}
Denote
\begin{equation}\label{xg}
\wh\bx_i=\left(\begin{array}{c}
\wh\bSigma_1^\top\\
\wh\bSigma_2^\top\\
\vdots \\
\wh\bSigma_r^\top
\end{array}\right)\be_i\quad\text{and}\quad\wh\bG=\left(\begin{array}{cc}
\wh\bSigma_1^\top&\wh\bSigma_0\\
\wh\bSigma_2^\top&\wh\bSigma_1^\top-\frac{1}{n}\by_{n-1}\by_{n}^\top\\
\vdots& \vdots\\
\wh\bSigma_r^\top&\wh\bSigma_{r-1}^\top-\frac{1}{n}\by_{n-r+1}\by_n^\top
\end{array}\right),
\end{equation}
where $\wh\bSigma_j=\frac{1}{n}\sum_{t=j+1}^n\by_{t}\by_{t-j}^\top$ for $j\geq1$. For technical convenience, we remove the last term of $\wh\bSigma_j^\top$ in the second half columns of $\wh\bG$ for $j\geq 1$.

By a similar argument as that in Section 2.2, we apply least squares method to estimate $(\bA,\bB)$ by solving the following minimization problems
\begin{equation}\label{m.m}
\mathop{\min}_{\btheta_i}\|\wh\bx_i-\wh\bG_i\btheta_i\|_2^2,\quad i=1,...,p,
\end{equation}
where $\btheta_i$ is a $\tau_i\times 1$ vector and $\wh\bG_i$ is the $rp\times \tau_i$ submatrix of $\wh\bG$ corresponding to the nonzero elements of $\ba_i$ and $\bb_i$. For each $i$, we denote $\hat{\hat{\bbeta}}_i$ the solution to the $i$-th equation of (\ref{m.m}). Then it follows from (\ref{m.m}) that
\begin{equation}\label{lse-m}
\hat{\hat{\bbeta}}_i = (\wh \bG_i^\top\wh \bG_i)^{-1} \wh \bG_i^\top \wh \bx_i, \quad i =1,..., p.
\end{equation}
Combining all the estimators in (\ref{eqn:YWest}) together leads to the
estimators for $\bA$ and $\bB$ which are denoted by, respectively, $\hat{\wh \bA}$ and $\hat{\wh \bB}$.

 Let $\mathbf{f}_{\epsilon_i}=(\frac{1}{n}\sum_{t=2}^n\by_{t-1}^\top\epsilon_{i,t},\frac{1}{n}\sum_{t=3}^n\by_{t-2}^\top\epsilon_{i,t},...,\frac{1}{n}\sum_{t=r+1}^n\by_{t-r}^\top\epsilon_{i,t})^\top$, it follows from (\ref{eqn:BSAR}) and (\ref{xg}) that
 \begin{equation}\label{xhat}
 \wh\bx_i=\wh\bG_i\bbeta_i+\mathbf{f}_{\epsilon_i}.
 \end{equation}
Hence it holds that
\begin{equation}\label{hhbeta}
\hat{\hat{\bbeta}}_i-\bbeta_i=(\wh \bG_i^\top\wh \bG_i)^{-1} \wh \bG_i^\top \mathbf{f}_{\epsilon_{i}},\quad i=1,...,p.
\end{equation}
We borrow the $\bu_{t,i}$ from Section 2.3, it is not hard to show that
\begin{equation}\label{hatg}
\wh\bG_i=\left(\begin{array}{c}
\frac{1}{n}\sum_{t=2}^n\by_{t-1}\bu_{t,i}^\top\\
\frac{1}{n}\sum_{t=3}^n\by_{t-2}\bu_{t,i}^\top\\
\vdots\\
\frac{1}{n}\sum_{t=r+1}^n\by_{t-r}\bu_{t,i}^\top
\end{array}\right).
\end{equation}
 We can define the corresponding residual sum of squares as
 (\ref{b7}) and estimate the bandwidth in the similar manner as in (\ref{eqn:ratio}).
From (\ref{hhbeta}) and (\ref{hatg}), we can see that, when $r=1$, the estimators in (\ref{hhbeta}) reduces to those in (\ref{eqn:YWest}).

% According to
% our limited simulation experience, it does not necessarily improve the
% estimation accuracy of $k_0$ with more estimation equations, we therefore
% omit the details here. In practice, with the estimator $\hat{k}_n$
% obtained by (\ref{eqn:ratio}), we can choose different $r$ to estimate
% $\bbeta_i$, simulation study suggests that $r=1$ is good enough to
% produce the estimators and  (\ref{hhbeta}) does not improve the
% estimation accuracy generally. See Tables \ref{tab:t1} and \ref{tab:t2}
% in Section 4 for details.

\section{ Theoretical properties}
\subsection{Notation and conditions}
We introduce some notations first. For a $p\times 1$ vector
$\bu=(u_1,..., u_p)^\top,$  $\|\bu\|_2 = (\sum_{i=1}^{p} u_i^2)^{1/2} $
is the Euclidean norm.  For a matrix $\bH=(h_{ij}),$  $\|\bH
\|_2=\sqrt{\lambda_{\max} (\bH^\top \bH ) }$ is the operator norm, where
$\lambda_{\max} (\cdot) $ denotes for the largest eigenvalue of a matrix.
We use $\lambda_{\min}(\cdot)$ to denote the smallest eigenvalue of a matrix.
For subset $S \subset \{1,...,p\}$, let $\bu_S=(u_j)_{j\in S}=(u_j,j \in
S)^\top$ be a column vector and $|S|$ be the cardinality of $S$. For a
matrix $\bSigma$, denote $\bSigma_S$ the sub-matrix consisting of the
columns of $\bSigma$ in $S$. A $p-$dimensional strictly stationary process $\by_t$ is
$\alpha$-mixing if
\begin{equation}
\label{eqn:alpha-mixing}
\alpha_p(k) \equiv \sup_{A \in \mathcal{F}_{-\infty}^0 , B \in \mathcal{F}_k^\infty} | P(A)P(B) - P(AB) | \rightarrow 0, \ \ \   \textnormal{as} \ k \rightarrow \infty,
\end{equation}
where $\mathcal{F}_i^j$ denotes the $\sigma$-algebra generated by $\{\my_t, i \leq t \leq j\}$.
We first introduce some regularity conditions.
\begin{itemize}
	\item[A1.] (i) The matrix $\bI_p -\bA$ is invertible, (ii) $\|(\mathbf{I}_p-\bA)^{-1}\bB\|_2<1$ and (iii) $\sum_{j=l}^\infty\|[(\bI_p-\bA)^{-1}\bB]^j\|_2\leq C_1 p^{-1/2}\rho^{l}$ for $l\geq 1$, some $\rho\in(0,1)$ and a positive constant $C_1$ independent of $p$.
	
	\item[A2.] \begin{itemize}
		\item[(a)] The innovations $\{\bve_t\}$ are independent and identically distributed (i.i.d.) satisfying $\cov(\by_{t-1},\bve_t)=0$, $\bxi_t:=(\bI_p-\bA)^{-1}\bve_t$ admits a density $g$ with $\int|g(\bv-\bu)-g(\bv)|d\bv<C_2\|\bu\|_2$ for $\bu\in R^p$, and $E\|\bxi_t\|_2^\delta<C_3p^{\delta/2}$ for some $\delta>0$, where $C_2$ and $C_3$ are positive constants independent of $p$.
		\item[(b)] The process $\by_t$ in model \eqref{eqn:BSAR}
is strictly stationary.% and $\alpha$-mixing with the mixing coefficients
%$\alpha(k)$, defined in \eqref{eqn:alpha-mixing}, satisfying the condition $
%\sum_{k=1}^\infty \alpha(k)^{\frac{\gamma}{4+\gamma}} < \infty  $ for
%some constant $\gamma >0$.	
		\item[(c)] For $\gamma >0$ specified in (b) above,
		\[ \sup_p E | \be_j^\top \bSigma_0 \by_t |^{4+\gamma} <\infty,
\ \ \sup_p E | \be_j^\top \bSigma_1 \by_t |^{4+\gamma} <\infty,  \quad
		  \sup_p E | \be_j^\top \by_t |^{4+\gamma} <\infty. \]
		The diagonal elements of $\bK_i$ defined in \eqref{eqn:Vi} are bounded uniformly in $p$.
	\end{itemize}	
	
\item[A3.]  The rank of $\bV_{i}$ is equal to $\tau_i$, where $\bV_i$ and $\tau_i$ are defined in (2.5) and (2.6), respectively.

\item[A4.] For any finite number of columns of $\bK_i$, denoted by $\bF_i$ and $\bH_i$ in matrix form and $\bF_i\neq\bH_i$, $\lambda_1\leq\lambda_{\min}\{\bF_i^\top(\bI_p-\bH_i(\bH_i^\top\bH_i)^{-1}\bH_i^\top)\bF_i\}\leq\lambda_{\max}\{\bF_i^\top(\bI_p-\bH_i(\bH_i^\top\bH_i)^{-1}\bH_i^\top)\bF_i\}\leq \lambda_2$ for some positive constants $\lambda_1\leq \lambda_2$.

\item[A5.] For each $i=1,..., p$, $|a_{i,i-k_0}|$ or $|a_{i,i+k_0}|$ as well as $|b_{i,i-k_0}|$ or $|b_{i,i+k_0}|$ is greater than $\{C_nk_0n^{-1}\log(p\vee n)\}^{1/2}$, where $C_n/n\rightarrow0$ and $C_n^2/(np)\rightarrow\infty$ as $n\rightarrow\infty$.

\item[A6.] $a_{i,j}$ and $b_{i,j}$ are bounded uniformly.% (Remark: A4, A5 and A6 are used to show the consistency of bandwidth selection.)
\end{itemize}

Conditions A1(i)-(ii) are standard for spatial econometric models, and
A1(iii) is for establishing the
$\alpha$-mixing condition in Lemma \ref{lm1} in the Appendix.
A sufficient condition for A1(iii) is
$\|(\bI_p-\bA)^{-1}\bB\|_2\leq Cp^{-1/2}\rho$ where $C$ is constant such
that $Cp^{-1/2}\rho<1$, and hence A1(ii) also holds.
Note that condition $p\to \infty$
is only a mathematical framework to reflect the scenarios when
the dimension $p$ is large (in relation to $n$), while in practice $p$
is always finite.
Therefore it makes sense to adopt the framework under which the limit
process of $\by_t$, as $p \to \infty$, is well-defined such that
$E\|\by_t\|_2<\infty$. This, therefore, implies that the
non-zero coefficients in $\bA$ and/or $\bB$ in model (\ref{eqn:BSAR})
decays to 0  as $p \to \infty$, which is reflected in
Condition A1(iii).  With this in mind, one can easily construct many
concrete examples fulfilling Condition A1(iii), including the models
with diagonal $\bA$ and $\bB$.
Condition A2(a) is for the validity of Lemmas 1 and 2 in
\cite{pham1985} in order to establish Lemma \ref{lm1} in the Appendix. Note that
$E\|\by_t\|_2<\infty$ implies that $E\|\bxi_t\|_2$  %and also $E\|\bve_t\|_2$
also remains finite as $p\to \infty$. Nevertheless
a large upper bound for $E\|\bxi_t\|_2^{\delta}$ in A2(a) is sufficient for
our analysis.  The strict stationarity in
Condition A2(b) is a non-asymptotic property,
i.e. for each $p$, we assume A2(b) holds. %It is also implied by A1(i)-(ii).
Similar to assumption A2(c) in Dou et al. (2016), Condition
A2(c) here limits the dependence across different spatial locations. It is
implied by, for example, the conditions imposed by \cite{yu2008quasi}.
Condition A2(c) can be verified under proper conditions with $\gamma=4$,
see Lemma 1 in Dou et al. (2016). Condition A3 ensures that $\bA$ and
$\bB$ are identifiable in (2.5). Conditions A4-A6 are imposed to prove
the consistency of our ratio estimator in (2.17). Condition A5 ensures
that the bandwidth is asymptotically identifiable, as $\{n^{-1}\log(p\vee
n)\}^{1/2}$ is the minimum order of a non-zero coefficient to be
identifiable, see, e.g., \cite{luo2013extended}. The proof of the
consistency can be simplified if the lower bound in A5 is replaced by
some positive constant, see the proof of Theorem 1 in the Appendix.

\subsection{Asymptotic  properties}

We first state the consistency of the ratio-based estimator $\wh k$
defined in (2.17), for determining the bandwidth parameter $k_0$.
\begin{theorem} \label{thm1}
Let Conditions A1-A6 hold and $p=o(n)$. Then $P(\wh k=k_0)\rightarrow1$, as $n\rightarrow\infty$.
\end{theorem}
\begin{remark}
In Theorem 1, $k_0$ is assumed to be fixed, as model (\ref{eqn:BSAR}) with
only small or moderately large $k_0$ are of practical usefulness.
Nevertheless Theorem~\ref{thm1} still holds
if $k_0$ diverges to $\infty$ together with $n,  p$, as long as $k_0<p$ and
 $k_0=o\{C_n^{-1}n/(\log (p\vee n))\}$, where $C_n$ is given in Condition A5.
See the proof of Theorem 1 in the Appendix.  \end{remark}

In the sequel $k_0$ is assumed to be either fixed or diverging with an appropriate rate.
% We will assume $k_0$ is fixed in the sequel for the purpose of stating
% the asymptotic properties of our estimators.
Since $k_0$ is unknown, we replace it by $\wh k$ in the estimation
procedure for $\bbeta_i$ described in Section 2, and still denote the
resulted estimators by $\wh \bbeta_i.$
For $i=1,...,p$, let
\begin{equation}\label{sigmae}
\bSigma_{\my,\bve_i}(j) = \textnormal{Cov}(\by_{t-1+j}\epsilon_{i,t+j}, \by_{t-1}\epsilon_{i,t}), \ j=0,1,2,...,
\end{equation}
\begin{equation}
\bSigma_{\my,\bve_i} = \bSigma_{\by,\bve_i}(0) + \sum_{j=1}^\infty \left[ \bSigma_{\my,\bve_i}(j)  + \bSigma^\top_{\by,\bve_i}(j) \right].
\end{equation}
Let $\bI_{S_i} = (\be_j,j\in S_i) \in \mR^{p\times |S_i|}$, $\bI_{S_i^+}=(\be_j,j\in S_i^+) \in \mR^{p\times |S_i^+|}$,
\begin{eqnarray}
\label{eqn:Vi}
\bK_i \equiv \left( \begin{array}{ll}
\bI_{S_i}^\top \bSigma_1 \bSigma_1^\top \bI_{S_i} & \bI_{S_i}^\top \bSigma_1 \bSigma_0 \bI_{S_i^+}  \\
\bI_{S_i^+}^\top \bSigma_0 \bSigma_1^\top \bI_{S_i} & \bI_{S_i^+}^\top \bSigma_0 \bSigma_0 \bI_{S_i^+}
\end{array}  \right)
\end{eqnarray}
and
\begin{eqnarray}
\label{eqn:Ui}
\bU_i \equiv \left( \begin{array}{ll}
\bI_{S_i}^\top \bSigma_1 \bSigma_{\my,\epsilon_i} \bSigma_1^\top \bI_{S_i} & \bI_{S_i}^\top \bSigma_1 \bSigma_{\my,\epsilon_i} \bSigma_0 \bI_{S_i^+}  \\
\bI_{S_i^+}^\top \bSigma_0 \bSigma_{\my,\epsilon_i} \bSigma_1^\top \bI_{S_i} & \bI_{S_i^+}^\top \bSigma_0 \bSigma_{\my,\epsilon_i} \bSigma_0 \bI_{S_i^+}
\end{array}  \right)
\end{eqnarray}

\begin{theorem}
	\label{theorem:asym-normal}
	Let Conditions A1-A6 hold. %Then the following assertions hold for
%$i=1, \cdots, p$, where $\tau_i$ is defined as in (\ref{b3}).
	\begin{itemize}
		\item[(i)] As $n\rightarrow \infty$, $p\rightarrow \infty$, and $p=o(\sqrt{n})$. { If $k_0$ is fixed, then}
		$$ \sqrt{n} \bU_i^{-1/2}\bK_i ( \wh \bbeta_i - \bbeta_i ) \rightarrow_d N(0,\bI_{\tau_i}),\quad i=1,...,p.  $$
		{If $k_0=o\{C_n^{-1}n/\log(p\vee n)\}$ and $\lambda_{\min}(\bK_i)\geq c>0$, then
		$$|| \wh \bbeta_i - \bbeta_i ||_2 = O_p(\sqrt{\frac{k_0}{n}}),\quad i=1,...,p.$$}
		\item[(ii)] As $n\rightarrow \infty$, $p\rightarrow \infty$, $\sqrt{n} = O(p)$, and $p=o(n)$. { If $k_0$ is fixed, then}
		$$ || \wh \bbeta_i - \bbeta_i ||_2 = O_p(\frac{p}{n}),\quad i=1,...,p.$$
		{If $k_0=o\{\min(C_n^{-1}n/\log(p\vee n),n/p)\}$ and $\lambda_{\min}(\bK_i)\geq c>0$, then
				$$ || \wh \bbeta_i - \bbeta_i ||_2 = O_p(k_0^{1/2}\frac{p}{n}),\quad i=1,...,p.$$}

\end{itemize}
\end{theorem}
\begin{remark}
If $p$ is fixed in theorem 2(i), the asymptotic normality can be rewritten as
$$\sqrt{n}( \wh \bbeta_i - \bbeta_i ) \rightarrow_d N(0,\bK_i^{-1}\bU_i\bK_i^{-1}),\quad i=1,...,p, $$
which achieves the standard $\sqrt{n}-$consistency. We also note that the convergence rate in Theorem \ref{theorem:asym-normal} is the same with that in Dou et al. (2016) when $k_0$ is fixed.
\end{remark}

%\begin{theorem}
%	\label{theorem:band}
%	Assume conditions A1-A6 hold. If $p=o(n)$, we have $P(\wt k = k_0) \rightarrow 1.$
%\end{theorem}
To derive the asymptotic properties of the estimators defined in (2.15), we introduce some new notations. For $i=1,...,p$, let
$$\bSigma_0^i=\Cov(\by_t,\bw_t^i),\quad \bSigma_1^i=\Cov(\by_t,\bw_{t-1}^i),$$
\begin{equation*}
\bSigma_{\bw^i,\bve_i}(j) = \textnormal{Cov}(\bw_{t-1+j}^i\epsilon_{i,t+j}, \bw_{t-1}^i\epsilon_{i,t}), \ j=0,1,2,...,
\end{equation*}
and
\begin{equation*}
\bSigma_{\bw^i,\bve_i} = \bSigma_{\bw^i,\bve_i}(0) + \sum_{j=1}^\infty \left[ \bSigma_{\bw^i,\bve_i}(j)  + \bSigma^\top_{\bw^i,\bve_i}(j) \right].
\end{equation*}
Let
\begin{eqnarray}
\label{eqn:ks}
\bK_i^* \equiv \left( \begin{array}{ll}
\bI_{S_i}^\top \bSigma_1^i (\bSigma_1^i)^\top \bI_{S_i} & \bI_{S_i}^\top \bSigma_1^i (\bSigma_0^i)^\top \bI_{S_i^+}  \\
\bI_{S_i^+}^\top \bSigma_0^i (\bSigma_1^i)^\top \bI_{S_i} & \bI_{S_i^+}^\top \bSigma_0^i (\bSigma_0^i)^\top \bI_{S_i^+}
\end{array}  \right)
\end{eqnarray}
and
\begin{eqnarray}
\label{eqn:us}
\bU_i^* \equiv \left( \begin{array}{ll}
\bI_{S_i}^\top \bSigma_1^i \bSigma_{\bw^i,\epsilon_i} (\bSigma_1^i)^\top \bI_{S_i} & \bI_{S_i}^\top \bSigma_1^i \bSigma_{\bw^i,\epsilon_i} (\bSigma_0^i)^\top \bI_{S_i^+}  \\
\bI_{S_i^+}^\top \bSigma_0^i \bSigma_{\bw^i,\epsilon_i} (\bSigma_1^i)^\top \bI_{S_i} & \bI_{S_i^+}^\top \bSigma_0^i \bSigma_{\bw^i,\epsilon_i} (\bSigma_0^i)^\top \bI_{S_i^+}
\end{array}  \right)
\end{eqnarray}
\begin{itemize}
\item[A7.](a) For $\gamma>0$ specified in A2(b),
\[ \sup_p E | \be_j^\top \bSigma_0^i \bw_t^i |^{4+\gamma} <\infty,
\ \ \sup_p E | \be_j^\top \bSigma_1^i \bw_t^i |^{4+\gamma} <\infty,  \quad
		  \sup_p E | \be_j^\top \by_t |^{4+\gamma} <\infty. \]
		The diagonal elements of $\bK_i^*$ defined in \eqref{eqn:ks} are bounded uniformly in $p$.\\(b) The rank of $\bW=E(\bw_{t-1}^i\bu_{t,i}^\top)$ is equal to $\tau_i$.
	
\end{itemize}
\begin{theorem}
Let Conditions A1, A2(a,b), and A3-A7 hold. As $n\rightarrow\infty$, $p\rightarrow\infty$ and $d_i=o(\sqrt{n})$, it holds for a fixed $k_0$ that
	$$ \sqrt{n} {\bU_i^*}^{-1/2}\bK_i^* ( \wt \bbeta_i - \bbeta_i ) \rightarrow_d N(0,\bI_{\tau_i}), \quad i=1,...,p,  $$
	where $\bK_i^*$ and $\bU_i^*$  are defined in (\ref{eqn:ks}) and (\ref{eqn:us}), respectively.
\end{theorem}
Theorem 3 indicates that the estimators defined in (2.15) are asymptotically normal with the standard rate as long as $d_i=o(\sqrt{n})$ and $k_0$ is fixed, and it does not impose any conditions directly on the size of $p$. {When $k_0$ is diverging, the convergence rate is the same as that in Theorem 2(i), and hence we omit the details here.}

To derive the asymptotic properties of the estimators $\hat{\hat{\bbeta}}_i$, similar to (\ref{sigmae})-(\ref{eqn:Ui}), let $\bQ_i$ be an $rp\times rp$ matrix which contains $r^2$ blocks with the $(j_1,j_2)-$th block
\begin{align}\label{Qe}
\bQ_i(j_1,j_2)=&\Cov(\by_{t-j_1}\epsilon_{i,t},\by_{t-j_2}\epsilon_{i,t})+\sum_{j=1}^\infty\{\Cov(\by_{t-j_1+j}\epsilon_{i,t+j},\by_{t-j_2}\epsilon_{i,t})\notag\\
&+\Cov(\by_{t-j_1}\epsilon_{i,t},\by_{t-j_2+j}\epsilon_{i,t+j})\}.
\end{align}
 We further define
 \begin{equation}
 \bR_i=\left(\begin{array}{cccc}
 \mathbf{I}_{S_i}^\top\bSigma_1&\mathbf{I}_{S_i}^\top\bSigma_2&\cdots&\mathbf{I}_{S_i}^\top\bSigma_r\\
  \mathbf{I}_{S_i^+}^\top\bSigma_0&\mathbf{I}_{S_i^+}^\top\bSigma_1&\cdots&\mathbf{I}_{S_i^+}^\top\bSigma_{r-1}
 \end{array}\right)
 \end{equation}
and
 \begin{equation}
 \bP_i=\left(\begin{array}{cc}
 \sum_{j=1}^r\mathbf{I}_{S_i}^\top\bSigma_j\bSigma_j^\top\mathbf{I}_{S_i}&\sum_{j=1}^r\mathbf{I}_{S_i}^\top\bSigma_j\bSigma_{j-1}^\top\mathbf{I}_{S_i^+}\\
  \sum_{j=1}^r\mathbf{I}_{S_i^+}^\top\bSigma_{j-1}\bSigma_j^\top\mathbf{I}_{S_i}&\sum_{j=1}^r\mathbf{I}_{S_i^+}^\top\bSigma_{j-1}\bSigma_{j-1}^\top\mathbf{I}_{S_i^+}
 \end{array}\right).
 \end{equation}
 By a similar proof as that of Theorem \ref{theorem:asym-normal}, we have the following theorem for the estimator $\wh{\wh{\bbeta}}_i$.

 \begin{theorem}
	\label{theorem:multiple}
	Let Conditions A1-A6 hold. %Then the following assertions hold for
%$i=1, \cdots, p$, where $\tau_i$ is defined as in (\ref{b3}).
	\begin{itemize}
		\item[(i)] As $n\rightarrow \infty$, $p\rightarrow \infty$, and $p=o(\sqrt{n})$. If $k_0$ is fixed, then
		$$ \sqrt{n} (\bR_i\bQ_i\bR_i^\top)^{-1/2}\bP_i ( \wh{\wh \bbeta}_i - \bbeta_i ) \rightarrow_d N(0,\bI_{\tau_i}),\quad i=1,...,p.  $$
		{If $k_0=o\{C_n^{-1}n/\log(p\vee n)\}$ and $\lambda_{\min}(\bP_i)\geq c>0$, then}
		$$||\wh {\wh \bbeta}_i - \bbeta_i ||_2 = O_p(\sqrt{\frac{k_0}{n}}),\quad i=1,...,p.$$
		\item[(ii)] As $n\rightarrow \infty$, $p\rightarrow \infty$, $\sqrt{n} = O(p)$, and $p=o(n)$. If $k_0$ is fixed, then
		$$ || \wh{\wh \bbeta}_i - \bbeta_i ||_2 = O_p(\frac{p}{n}),\quad i=1,...,p.$$
		{If $k_0=o\{\min(C_n^{-1}n/\log(p\vee n),n/p)\}$ and $\lambda_{\min}(\bP_i)\geq c>0$, then}
				$$ ||\wh {\wh \bbeta}_i - \bbeta_i ||_2 = O_p(k_0^{1/2}\frac{p}{n}),\quad i=1,...,p.$$
\end{itemize}
\end{theorem}
 \begin{remark}
 If we compare the results in Theorem \ref{theorem:multiple} with those in Theorem \ref{theorem:asym-normal}, we can see that, given a finite positive integer $r$, the rates of the estimation errors are the same. When $p$ is fixed, we can also achieve the standard $\sqrt{n}-$consistency in Theorem \ref{theorem:multiple} with the covariance $\bP_i^{-1}(\bR_i\bQ_i\bR_i^\top)\bP_i^{-1}$, which is different from that in Theorem \ref{theorem:asym-normal}. Our simulation results in Tables \ref{tab:t1} and \ref{tab:t2} suggest that $r=1$ is good enough to produce the estimators with smaller estimation errors.
 \end{remark}

\section{Numerical properties}

\subsection{Simulation}
To evaluate the finite sample performance of our proposed
method, we  conduct simulations as follows.
We simulate $\by_t$ from model (\ref{b2}) with independent
and $N(0, 1)$ innovations $\ve_{i,t}$. We consider two settings for
coefficient matrices $\bA=(a_{i,j})$ and $\bB=(b_{i,j})$.

\noindent
\textbf{Case 1}.
Elements $a_{i,j}, b_{i,j}$ for $|i-j|=k_0$ are drawn independently from
uniform distribution on two points $\{ -2, 2\}$, and $a_{i,j}$ for $0<
|i-j|< k_0$ and $b_{i,j}$ for $|i-j|< k_0$ are drawn independently
from  the mixture distribution $\omega I_{\{0 \}} + (1-\omega) N(0, 1)$
with $P(\omega=1) =0.4 = 1 - P(\omega=0)$. We then rescale  $\bA$ and $\bB$ to  $\eta_1 \cdot \bA/\|\bA\|_2$ and $\eta_2 \cdot \bB/\|\bB\|_2$,
where $\eta_1$ and $\eta_2$  are drawn independently from $U[0.4, 0.8]$.

\noindent
\textbf{Case 2}. Elements $a_{i,j}, b_{i,j}$ for $|i-j|=k_0$ are drawn independently from
$U([-2.5, -1.5]\cup [1.5,2.5])$,  and $a_{i,j}$ for $0< |i-j|< k_0$ and $b_{i,j}$ for
$|i-j|< k_0$ are drawn independently
from $U[-1, 1]$.  We then rescale $\bA$ and $\bB$ as in Case 1 above.

For each model, we set sample size $n=500, 1,000$, and $2,000$ and
dimension of time series
$p=100, 300, 500, 800$ and $1,000$. This leads to the 15 different  $(n, p)$ combinations.
For each setting, we replicate the experiment 500 times, and calculate the relative
frequencies (\%) for the occurrence of events $\{\wh k = k_0\}$, $\{\wh k > k_0\}$ and $\{\wh k < k_0\}$
in the 500 replications. We also calculate the means and the standard deviations of the estimation
errors $\| \bA - \wh \bA\|_2$ and $\| \bB - \wh \bB\|_2$.
{The results with the bandwidth parameter $k_0=3$,  the upper bound
$K=10$ in (\ref{eqn:ratio}), and  $ r=1,2,3$ in (\ref{myk}), are reported in Tables \ref{tab:t1} and \ref{tab:t2}.
For each setting, we also report the signal-to-noise ratio defined as
\[
{\rm SNR}= {\rm tr}\{\var(\by_t)\}\big/ {\rm tr}\{(\bI_p-\bA)^{-1}\var(\bve_t)(\bI_p-\bA^\top)^{-1}\}.
\]
As indicated clearly in Tables  \ref{tab:t1} and \ref{tab:t2}, when the sample size $n$ increases, the errors in estimating the coefficient matrices $\bA$ and $\bB$
decrease while the relative frequencies (\%) for the correct specification of the bandwidth parameter $k_0$
 increase. Note that the errors in estimating $\bA$  based on $r=1,2, 3$ show no clear difference. However, when $n$ and $p$  are fixed, the errors in estimating $\bB$ are increasing  with $r$. This suggests that $r=1$ is good enough.  {We also notice that when $p$ is fixed, the standard deviations of $\|\bA-\hat\bA\|$ and  $\|\bB-\hat\bB\|$ are not necessarily decreasing with $n$, see, for example, $p=100$ and $1,000$ in Table 2. This is affected by the fluctuations of $\wh k$ and a dominant proportion of either $\{\wh k=k_0\}$ or $\{\wh k>k_0\}$ usually produces more stable estimation errors.}
%Note that the errors in estimating $\bA$ are greater than
%those in estimating $\bB$, indicating $\bA$ is more difficult to estimate
%than $\bB$. This is reasonable as $\bB$
%occurs only in the conditional mean while $\bA$ also occurs in the variance in model (\ref{b2}).
Moreover, there is no clear pattern in performance with respect to different values
of the dimension $p$. This is  due to  the fact that the signal-to-noise
ratio does not vary monotonically with respect to $p$. {Overall,} the larger the signal-to-noise
ratio is, the better performance is observed in estimating both the
coefficient matrices $(\bA, \bB)$ and the bandwidth parameter $k_0$; see Tables \ref{tab:t1} and \ref{tab:t2}.
The results with different values of $k_0$ and $K$ are similar, and therefore omitted to save the space.

{ To compare the estimators in (\ref{eqn:YWest}) and (\ref{b11}), we generate the data as Case 2 with $K=5$ and $k_0=1$. For each $p=50$, $75$, $100$ and $125$, we set the sample size $n=2,500$, $5,000$ and $10,000$, respectively. In addition, we choose $d_i=\min(p,\lbrack n^{0.495}\rbrack)$ and denote the two estimators by Estimate I and Estimator II, respectively. The proportions of $ \{\hat k=k_0\}$, $ \{\hat k>k_0\}$  and $ \{\hat k<k_0\}$ based on $r=1$, the mean and standard deviations of $ \|\bA-\hat \bA\|_2 $
and $ \|\bB-\hat \bB\|_2$ are reported in Table 3. We can see from Table 3 that for each $p$, the estimation errors decrease as the sample size increases. On the other hand, for each $p$ and $n$, the root-$n$ consistent estimator (Estimator II) tends to have larger estimation errors. This is also confirmed by the simulation results in Dou et al. (2016) since (\ref{b11}) only makes use of part of the information  for the parameters as long as  $d_i<p$. }

{The comparisons of our method to those in Dou et al. (2016) and Yu et al. (2008) are studied in a supplementary material in order to save space.}
\subsection{Illustration with real data}
We illustrate the proposed model with  two real data sets in this section.

\noindent
\textbf{Example 1}.
With the rapid economic growth in China in recent years, there has also been
 a substantial increase in energy consumption, leading to
serious air pollution in large part of China \citep{Wang2002PM2.5,Wang2015PM2.5}. One of the important pollution indicators is the so-called $\mbox{PM}_{2.5}$
index, which measures the concentration level of  fine particulate matter
in the air. The $\mbox{PM}_{2.5}$ pollution is severe in the north
China plain (i.e., Beijing, Tianjin, and  Hebei province).
We consider here the  hourly  $\mbox{PM}_{2.5}$ readings  at the 36 monitoring stations  in
Beijing area in the period  of 1 April --- 30 June 2016 (i.e., $n = 2184, p = 36$).
Fig.\ref{fig1} is the map of those 36 stations.
  Fig.\ref{fig2} displays the original hourly $\mbox{PM}_{2.5}$ records from  three
randomly selected stations (i.e.,  Miyun, Huairou, and Shunyi).
We apply the logarithmic transformation to the data and substract the mean for each of the 36
transformed series.
 Fig.\ref{fig3} plots the three transformed series from those in Fig.\ref{fig2}.
To fit model (\ref{eqn:BSAR}) to the transformed data,
the 36 monitoring stations need to be arranged in a unilateral order.
We consider the five possible options for the ordering, i.e., we order the
stations along the directions from
north to south, from west to east,  from northwest  to southeast,  from
northeast to southwest,  and we  also order the stations according to their geographic distances to
Miyun -- a station at the northeast corner of the region; see Fig.\ref{fig1}.
We select an ordering, among those five,
according to a version of moving-window cross validation method; see below.

For each given ordering, we apply the ratio-based method to estimate the bandwidth parameter $k_0$.
We apply a moving-window cross-validation scheme to calculate the post-sample predictive
errors, i.e. for each of $t=2001, \cdots, 2184$, we fit a model using
only its 2000 immediate past observations.  We then calculate one-step ahead and
two-step ahead predictive errors.
The results are summarized in Table~\ref{t3}. Based on both the
one-step ahead and  two-step ahead mean squared   predictive errors, the ordering
from west to east is preferred
with the ordering from  north to south as the close second. Note that for both of the orderings,
the estimated bandwidth parameter is $\wh k =5$.

According to the Air Quality Standard in China, the $\mbox{PM}_{2.5}$ pollution is marked
at 7 different levels: Level 1 indicates the lowest pollution with the
$\mbox{PM}_{2.5}$ concentration below 35 micrograms per cubic meter of
air, and Level 7 corresponds to the worst scenarios with the $\mbox{PM}_{2.5}$
concentration exceeding 500 micrograms per cubic meter of air.
For general public the prediction for the pollution level is of more interest
than that for a concrete concentration value. Table~\ref{t4} presents the percentages of
the corrected one-step ahead and two-step ahead (post-sample) predictions at each of
the 7 levels based on the five different orderings. It is easy to see from Table~\ref{t4}
that the higher the pollution level is, the more accurate the prediction is.
Especially Level 6 and 7 pollution can always be correctly predicted based on all the
five models. The preferred models with the ordering from north to south or from west to east
provide overall higher percentages of correct prediction across the 7 pollution levels
than the other three models.

% In realistic, people usually use the $\mbox{PM}_{2.5}$ data to do air
% quality forecast and based on China air quality standards, the air
% quality can be classified into 7 different levels. The higher the level
% is, the worse of the air quality. Thus, Level 1 corresponds to the best
% air quality with $\mbox{PM}_{2.5}$ values less than 35 micrograms per
% cubic meter of air, and the extremely polluted situation corresponds to
% level 7   with $\mbox{PM}_{2.5}$ values more than 500.
% Table~\ref{t4} has summarized the detailed prediction accuracy rate for different
%  air pollution levels over 36 stations. The accuracy rate is summarized in percentage. For instance, based on ordering from north to south, with the chance of $68.4\%$,  our model can correctly predict the  one- step ahead air quality  as level 1.
%  As we can see from this table, our method can always have very high  prediction accuracy for extremely poor air quality (level 5, 6, and 7). The high prediction accuracy can help us to get better preparation for the air pollution. Moreover,  the banded
% structure can help us to understand how the air pollution is related with its neighborhood.

\noindent
\textbf{Example 2}.
 Now we consider the annual mortality rates in the period of
1872 --- 2009 for the Italian population at age $i$,
for $i=10, 11, \cdots, 50$. The data were
downloaded from http://www.mortality.org/.
  Let $m_{i,t}$ be the  original mortality rate (male and female in
total) at age $i$ in the $t$-th year.
  Fig.\ref{fig4} displays the three series of $m_{i,t}$  with age $i=10, 30,$ and $50$ respectively.
 Overall the mortality rates
decrease for all age groups  over the years except in the period of
World War I in 1914 -- 1918 and World War II in 1939 -- 1945.
Let $\{y_{i,t}, t=1872,\cdots, 2009\}$ be the centered log-scaled mortality rates for
the $i$-th age group, $i=10, 11, \cdots, 50$. Thus $p=41$ and $n=138$.
This orders the components of $\by_t$ naturally by the age.
The ratio-based method leads to the estimated bandwidth parameter $\wh k =1$ for
this data set. We compute both one-step ahead and two-step ahead
post-sample predictive errors for the last 8 data points for each of 41
series. The results are reported in Table~\ref{tab:t5}.

Also included in Table~\ref{tab:t5} are the predictive errors based on
the spatio-temporal  model of \cite{dou2016generalized} which uses a
known spatial weight matrix
but with different scalar parameters for different location.  The spatial
weight matrix is defined as $W=(w_{i,j})$ with $w_{i,j}=a_{i,j}/\sum_{i}a_{i,j}$ for $i \neq j$,  and 0 for $i=j$. We use two specifications for $a_{i,j}$: (i) a distance measure
$a_{i,j}=(1+|i-j|)^{-1}$, and (ii) a correlation measure with $a_{i,j}$  taken as
the absolute sample correlation between $y_{i,t}$ and $y_{j,t}$.
Table~\ref{tab:t5} indicates clearly that the proposed banded model performs
better than \cite{dou2016generalized}'s model in post-sample forecasting.

\section{Concluding remarks}

We propose in this paper a new class of banded spatio-temporal models. The setting does not
require pre-specified spatial weight matrices. The coefficient matrices are estimated
by a generalized method of moments estimation based on a Yule-Walker equation. The bandwidth
of the coefficient matrices is determined by a ratio-based method.

%%%%%%%%%%%%%%%%%%%%%%%%%%%%%%%%%%%%%%%%%
\noindent {\large\bf Acknowledgments}

We are grateful to the  Editors  and the anonymous referees for their insightful comments and suggestions that have substantially improved the presentation and the content of this paper.
We also acknowledge the partial support of China's National Key Research
Special Program Grant  2016YFC0207702, National Natural Science Foundation of
China (NSFC, 71532001, 11525101), Science Foundation of Ministry of Education  of
China  17YJC910006, and the UK EPSRC research grant EP/L01226X/1.

\section*{Appendix: Proofs}
We present the proofs for Theorem 1 and Theorem 2 in this appendix. The idea of the proof for Theorem 2 is similar to that in Dou et al. (2016), but our setting is different since we have a banded structure and the convergence is a multivariate case. The proof for Theorem 4 follows directly from that of Theorem 2 and the proof for Theorem 3 is similar and simpler than that of Theorem 2,  and they are therefore omitted.  We use $C$ to denote a generic positive constant, which may be different at different places.
\renewcommand{\theequation}{A.\arabic{equation}}
\setcounter{equation}{0}
%\subsection{Appendix C. Proof of Theorem \ref{theorem:band}}
%\renewcommand{\theequation}{A.\arabic{equation}}
%\setcounter{equation}{0}

Before we prove the main theorems for the estimators in Section 2, we first give a lemma showing that the process $\{\by_t\}$ is strongly mixing under some regularity conditions.
\begin{lemma}\label{lm1}
If Conditions A1 and A2(a) hold, the process $\by_t$ is $\alpha-$mixing with the mixing coefficients $\alpha_p(k)$, defined in (\ref{eqn:alpha-mixing}), satisfying the condition $\sum_{k=1}^\infty \alpha_p(k)^{\frac{\gamma}{4+\gamma}} < \infty  $ uniformly for
 all sufficiently large $p$ and some constant $\gamma >0$.	
\end{lemma}
{\bf Proof:} It suffices to show that, uniformly for sufficiently large $p$, $\alpha_p(k)=O(a^k)$ for $k\geq 1$ and some constant $a\in(0,1)$. Let
$$\bD=(\bI_p-\bA)^{-1}\bB\,\,\text{and}\,\,\bxi_t=(\bI_p-\bA)^{-1}\bve_t,$$
where $\bxi_t$ is the same with that in Condition A2. It follows from (\ref{b2}) and Condition A1 that
\begin{equation} \label{bb2}
 \my_t = \bD\my_{t-1} + \bxi_t=\sum_{k=0}^\infty \bD^k\bxi_{t-k},
\end{equation}
where $\bD^0=\bI_p$. Note that the results of Lemmas 2.1-2.2 in \cite{pham1985} are still valid for model (\ref{bb2}) under assumptions A1-A2(a). To avoid the confusion of the notation $\alpha(j)$ in \cite{pham1985}, here we define $\sigma(j)=\sum_{k\geq j}\|\bD^k\|_2$ to replace the expression  of $\alpha(j)$ in their paper. By Lemmas 2.1-2.2 and the proof of Theorem 2.1 therein,  we have
 $$\|\Delta_n\|_{L^1}\leq C\sum_{j=n}^\infty\sigma(j)c_j+2\sum_{j=n}^\infty P(\|\bxi_t\|_2>c_j),$$
where $\Delta_n(x)$ is defined as (1.1) in Pham and Tran (1985), $\|\Delta_n\|_{L^1}$ is the $L^1$-norm of $\Delta_n(x)$ and $C$ is a generic constant independent of $p$.
 Let $c_j=p^{\delta/[2(1+\delta)]}\sigma(j)^{-1/(1+\delta)}$, by assumptions A1-A2(a) and Schwartz inequality, we have
 $P(\|\bxi_t\|_2>c_j)\leq E\|\bxi_t\|_2^\delta/c_j^\delta$ and hence
  $$\|\Delta_n\|_{L^1}\leq C\sum_{j=n}^\infty[\sum_{i=j}^\infty\rho^i]^{\delta/(1+\delta)}=O([\rho^{\delta/(1+\delta)}]^n)=O(a^n),$$
  where $a=\rho^{\delta/(1+\delta)}$.
The conclusion of Lemma \ref{lm1} follows from the fact that  $\alpha_p(n)\leq 4\|\Delta_n\|_{L^1}$, see Pham and Tran (1985) for details. This completes the proof. $\Box$

\noindent
{\bf Proof of Theorem 1.} For each $i=1,...,p$, let $\wh k_i=\mathop{\arg\max}_{1<k\leq K}( \RSS_i(k-1) + w_n )\big/  (\RSS_i(k) +  w_n)$. Our goal is to prove that $P(\wh k = k_0) \rightarrow 1.$ It is sufficient to show that
\begin{equation}\label{A.1}
 P(\wh k < k_0) \rightarrow 0  \quad \textnormal{and}  \quad P(\wh k > k_0) \rightarrow 0,
\end{equation}
respectively. We first investigate the convergence rate of $\RSS_i(k)$, which is crucial for proving the statement (\ref{A.1}) above. For $k\geq k_0$, let
$$\wh \bV_{i,k}=(\bS_{i,k}^{(1)},\wh\bSigma^\top_{1,k_0},\bS_{i,k}^{(2)},\bS_{i,k}^{(3)},\wh\bSigma_{0,k_0},\bS_{i,k}^{(4)}),\quad \bbeta_{i,k}=(\ba_{i,k}^{(1)^\top},\ba_{i,k_0}^\top,\ba_{i,k}^{(2)^\top},\bb_{i,k}^{(1)^\top},\bb_{i,k_0}^\top,\bb_{i,k}^{(2)^\top})^\top,$$
where $\wh \bV_{i,k_0}=\wh\bV_i=(\wh\bSigma^\top_{1,k_0},\wh\bSigma_{0,k_0})$ and $\bbeta_{i,k_0}=\bbeta_{i}=(\ba_{i,k_0}^\top,\bb_{i,k_0}^\top)^\top$, which correspond to the $\tau_i$ columns of $(\wh\bSigma_1^\top,\wh\bSigma_0)$ and $\tau_i$ non-zero elements of $(\ba_i^\top,\bb_{i}^\top)^\top$, respectively. Define $\bH_{i,k}=\wh\bV_{ik}(\wh\bV_{i,k}^\top\wh\bV_{i,k})^{-1}\wh\bV_{i,k}^\top$, it follows from (\ref{b7}) and (\ref{b13}) that
\begin{equation}
\RSS_i(k_0)=\frac{1}{p}\|(\bI-\bH_{i,k_0})\frac{1}{n}\sum_{t=1}^n\by_{t-1}\epsilon_{i,t}\|_2^2\leq \frac{1}{p}\|\bI-\bH_{i,k_0}\|_2^2\|\frac{1}{n}\sum_{t=1}^n\by_{t-1}\epsilon_{i,t}\|_2^2.
\end{equation}
Since $(\bI-\bH_{i,k_0})^2=\bI-\bH_{i,k_0}$ is a projection matrix, we have $\|\bI-\bH_{i,k_0}\|_2^2\leq 1$. Then, by a similar argument as (14) in Dou et al. (2016) or (\ref{rates}) below in the proof of Theorem 2, we conclude that
\begin{equation}\label{A.3}
\RSS_i(k_0)\leq \frac{1}{p}\|\frac{1}{n}\sum_{t=1}^n\by_{t-1}\epsilon_{i,t}\|_2^2=O_p(\frac{1}{n}).
\end{equation}
When $k>k_0$, (\ref{b7}) can be rewritten as
$$\RSS_i(k)=\frac{1}{p}\min_{\bv_1,\bv_2}\|\wh\bz_i-\wh\bV_{i,k_0}\bv_1-\bS_{i,k}\bv_2\|_2^2,$$
where $\bS_{i,k}=(\bS_{i,k}^{(1)},\bS_{i,k}^{(2)},\bS_{i,k}^{(3)},\bS_{i,k}^{(4)})$. Let $\wt\bS_{i,k}=(\bI_p-\bH_{i,k_0})\bS_{i,k}$, it can be verified that
$$\RSS_i(k)=\frac{1}{p}\|(\bI_p-\bH_{i,k_0})\wh\bz_{i}\|_2^2-\frac{1}{p}\|\wt\bS_{i,k}\wh\bv_2\|_2^2=\RSS_i(k_0)-\frac{1}{p}\|\wt\bS_{i,k}\wh\bv_2\|_2^2,$$
where $\wh\bv_2=(\wt\bS_{i,k}^\top\wt\bS_{i,k})^{-1}\wt\bS_{i,k}^\top\wh\bz_i$. By (\ref{b13}) and (\ref{A.3}), we have
\begin{align}\label{A.4}
\RSS_i(k)=&\RSS_i(k_0)-\frac{1}{p}\|\wt\bS_{i,k}(\wt\bS_{i,k}^\top\wt\bS_{i,k})^{-1}\wt\bS_{i,k}^\top\frac{1}{n}\sum_{t=1}^n\by_{t-1}\epsilon_{i,t}\|_2^2\notag\\
\leq&O_p(\frac{1}{n})+\frac{1}{p}\|\wt\bS_{i,k}(\wt\bS_{i,k}^\top\wt\bS_{i,k})^{-1}\wt\bS_{i,k}^\top\|_2^2\|\frac{1}{n}\sum_{t=1}^n\by_{t-1}\epsilon_{i,t}\|_2^2\notag\\
=&O_p(\frac{1}{n}),
\end{align}
since $\wt\bS_{i,k}(\wt\bS_{i,k}^\top\wt\bS_{i,k})^{-1}\wt\bS_{i,k}^\top$ is a projection matrix.

Similarly, for $k<k_0$, we define
$$\wh \bV_{i,k_0}=(\bJ_{i,k}^{(1)},\wh\bSigma^\top_{1,k},\bJ_{i,k}^{(2)},\bJ_{i,k}^{(3)},\wh\bSigma_{0,k_0},\bJ_{i,k}^{(4)}),\quad \bbeta_{i,k_0}=(\mathbf{c}_{i,k}^{(1)^\top},\ba_{i,k}^\top,\mathbf{c}_{i,k}^{(2)^\top},\mathbf{d}_{i,k}^{(1)^\top},\bb_{i,k}^\top,\mathbf{d}_{i,k}^{(2)^\top})^\top,$$
where $\wh\bV_{i,k}=(\wh\bSigma^\top_{1,k},\wh\bSigma_{0,k})$ and $\bbeta_{i,k=}(\ba_{i,k}^\top,\bb_{i,k}^\top)^\top$, which correspond to $\tau_i(k)$ columns of $(\wh\bSigma_1^\top,\bSigma_0)$ and $\tau_i(k)$ elements of $\beta_i$. $\tau_i(k)$ is defined as (2.6) with $k_0$ replaced by $k$. It follows from (\ref{b13}) that
\begin{equation}\label{A.5}
\wh\bz_i=\wh\bV_{i,k_0}\bbeta_{i,k_0}+\frac{1}{n}\sum_{t=1}^n\by_{t-1}\epsilon_{i,t}
=\wh\bV_{i,k}\bbeta_{i,k}+\bJ_{i,k}\mathbf{\delta}_{i,k}+\frac{1}{n}\sum_{t=1}^n\by_{t-1}\epsilon_{i,t},
\end{equation}
where $\bJ_{i,k}=(\bJ_{i,k}^{(1)},\bJ_{i,k}^{(2)},\bJ_{i,k}^{(3)},\bJ_{i,k}^{(4)})$ and $\mathbf{\delta}_{i,k}=(\mathbf{c}_{i,k}^{(1)^\top},\mathbf{c}_{i,k}^{(2)^\top},\mathbf{d}_{i,k}^{(1)^\top},\mathbf{d}_{i,k}^{(2)^\top})^\top$. By (\ref{b7}) and (\ref{A.5}),
\begin{align}\label{A.6}
\RSS_i(k)=&\frac{1}{p}\|(\bI_p-\bH_{i,k})\wh\bz_i\|_2^2\notag\\
=&\frac{1}{p}\|(\bI-\bH_{i,k})\bJ_{i,k}\mathbf{\delta}_{i,k}+(\bI-\bH_{i,k})\frac{1}{n}\sum_{t=1}^n\by_{t-1}\epsilon_{i,t}\|_2^2\notag\\
=&\frac{1}{p}\|(\bI-\bH_{i,k})\bJ_{i,k}\mathbf{\delta}_{i,k}\|_2^2+\frac{1}{p}\|(\bI-\bH_{i,k})\frac{1}{n}\sum_{t=1}^n\by_{t-1}\epsilon_{i,t}\|_2^2\notag\\
&+\frac{2}{p}\mathbf{\delta}_{i,k}^\top\bJ_{i,k}^\top(\bI-\bH_{i,k})\frac{1}{n}\sum_{t=1}^n\by_{t-1}\epsilon_{i,t}.
\end{align}
By Condition A5, we have
$$\lambda_{\min}\{\bJ_{i,k}^\top(\bI_p-\bH_{i,k})\bJ_{i,k}\}\geq\lambda_1\quad\text{and}\quad\lambda_{\max}\{\bJ_{i,k}^\top(\bI_p-\bH_{i,k})\bJ_{i,k}\}\leq \lambda_2.$$
Then, the first term of (\ref{A.6}) can be bounded by
\begin{equation}\label{A.7}
\frac{\lambda_1}{p}(a_{i,i-k_0}^2+a_{i,i+k_0}^2+b_{i,i-k_0}^2+b_{i,i+k_0}^2)\leq \frac{1}{p}\|(\bI-\bH_{i,k})\bJ_{i,k}\mathbf{\delta}_{i,k}\|_2^2\leq \frac{\lambda_2}{p}\|\bbeta_{i,k_0}\|_2^2.
\end{equation}
By Conditions A6 and A7, (\ref{A.7}) can be relaxed to
\begin{equation}
\frac{C_nk_0\lambda_1\log(p\vee n)}{np}\leq\frac{1}{p}\|(\bI-\bH_{i,k})\bJ_{i,k}\mathbf{\delta}_{i,k}\|_2^2\leq\frac{O(1)\lambda_2}{p}.
\end{equation}
The second term is of order $O_p(\frac{1}{n})$ by (\ref{A.3}). By Cauchy-Schwarz inequality, the third term can be bounded by the sum of the first and the second terms. As a result,
\begin{equation}\label{A.9}
\frac{C_nk_0\lambda_1\log(p\vee n)}{np}+O_p(\frac{1}{n})\leq \RSS_i(k)\leq \frac{O(1)\lambda_2}{p}.
\end{equation}
Now we are able to prove (\ref{A.1}). To prove $P(\wh k>k_0)\rightarrow0$, we note that $P(\wh k>k_0)\leq P(\wh k_i>k_0)$ for some $i\in\{1,...,p\}$ and the event $\{\wh k_i>k_0\}$ implies
$$A_{in}\equiv\{\max\limits_{k>k_0}\frac{\RSS_i(k-1)+w_n}{\RSS_i(k)+w_n}>\frac{\RSS_i(k_0-1)+w_n}{\RSS_i(k_0)+w_n}\}.$$
Then, we only need to show that $P(A_{in})\rightarrow0$ for some $i$. By (\ref{A.4}), (\ref{A.9}) and Condition A5,
\begin{equation}\label{A.10}
\frac{\RSS_i(k_0-1)+w_n}{\RSS_i(k_0)+w_n}\geq \frac{\lambda_1 C_n k_0 \log(p\vee n)/(np)}{O_p(1/n)}\rightarrow\infty,
\end{equation}
and
\begin{equation}\label{A.11}
\max\limits_{k>k_0}\frac{\RSS_i(k-1)+w_n}{\RSS_i(k)+w_n}\leq \frac{w_n+O_p(1/n)}{w_n}=O_p(1).
\end{equation}
It follows from (\ref{A.10}) and (\ref{A.11}) that $P(A_{in})\rightarrow0$, and hence $P(\wh k>k_0)\rightarrow0$.

Similarly, to prove $P(\wh k<k_0)\rightarrow0$, we only need to show that $P(B_{in})\rightarrow0$ for some $i\in\{1,...,p\}$, where
$$B_{in}\equiv \{\max\limits_{k<k_0}\frac{\RSS_i(k-1)+w_n}{\RSS_i(k)+w_n}>\frac{\RSS_i(k_0-1)+w_n}{\RSS_i(k_0)+w_n}\}.$$
By (\ref{A.9}),

\begin{equation}\label{A.12}
\max\limits_{k<k_0}\frac{\RSS_i(k-1)+w_n}{\RSS_i(k)+w_n}\leq \frac{w_n+O_p(1)\lambda_2/p}{{C_nk_0\lambda_1\log(p\vee n)/(np)}+O_p(1/n)}.
\end{equation}
We now compare the ratio between the upper bound of (\ref{A.12}) and the lower bound of (\ref{A.10}),
\begin{align}\label{A.13}
&\left\{\frac{w_n+O_p(1)\lambda_2/p}{{C_nk_0\lambda_1\log(p\vee n)/(np)}+O_p(1/n)}\right\}\bigg/\left\{\frac{\lambda_1 C_n k_0 \log(p\vee n)/(np)}{O_p(1/n)}\right\}\notag\\
&=\frac{O_p(p^2)+O_p(np)}{O_p(C_n^2(\log(p\vee n))^2)+O_p(pC_n\log(p\vee n))}\rightarrow0,
\end{align}
as long as $C_n^2/(np)\rightarrow\infty$. It follows from (\ref{A.10}), (\ref{A.12}) and (\ref{A.13}) that $P(B_{in})\rightarrow0$. If $k_0$ is not fixed, the upper bound in (\ref{A.9}) can be replaced by $O(1)k_0\lambda_2/p$, (\ref{A.13}) still holds under Conditions A4-A6. This completes the proof of Theorem 1. $\Box$

\noindent
%\subsection{Appendix A. Proof of Proposition 1.}
{\bf Proof of Theorem 2.}
By Theorem 1, with probability tending to one, $\wh k=k_0$, and thus it suffices to consider the set $\mathcal{A}_n=\{\wh k=k_0\}$.  Over the set $\mathcal{A}_n$, to prove part $(i)$ of Theorem 2 for a fixed $k_0$,  following the same arguments in Dou et al. (2016), we only need to verify the assertions (1) and (2) below.
	\begin{itemize}
		\item[(1)]
		\begin{eqnarray}
		\sqrt{n} \bU_i^{-\frac{1}{2}} \wh \bV_i^\top  \left( \frac{1}{n} \sum_{t=2}^{n}  \by_{t-1} \epsilon_{i,t} \right) &&= \sqrt{n} \bU_i^{-\frac{1}{2}} \left(
		\begin{array}{l}
		\frac{1}{n} \sum_{t=2}^{n}  (y_{j,t})_{j \in S_i}\by^\top_{t-1}  \left( \frac{1}{n} \sum_{t=2}^{n}  \by_{t-1} \epsilon_{i,t} \right) \\
		\frac{1}{n} \sum_{t=2}^{n}  (y_{j,t-1})_{j \in S_i^+}\by^\top_{t-1}   \left( \frac{1}{n} \sum_{t=2}^{n}  \by_{t-1} \epsilon_{i,t} \right)
		\end{array}
		\right)  \nonumber \\
		 &&\rightarrow_d N(0,\bI_{\tau_i}). \nonumber
		\end{eqnarray}		
		\item[(2)] $\bK_i (\wh \bV_i^\top \wh\bV_i)^{-1} \rightarrow_p \bI_{\tau_i}.$
	\end{itemize}
	To prove assertion (1), it suffices to show that for any nonzero vector $\vu = (\vu_1^\top, \vu_2^\top)^\top \in \mR^{\tau_i}$, where $\vu_1 \in \mR^{S_i}$, $\vu_2 \in \mR^{S_i^+}$ and $\tau_i = |S_i| + |S_i^+|$, the linear
	combination
	\begin{eqnarray}
\label{A.14} \sqrt{n} \bu^\top\left(
		\begin{array}{l}
		\frac{1}{n} \sum_{t=2}^{n}  (y_{j,t})_{j \in S_i}\by^\top_{t-1}  \left( \frac{1}{n} \sum_{t=2}^{n}  \by_{t-1} \epsilon_{i,t} \right) \\
		\frac{1}{n} \sum_{t=2}^{n}  (y_{j,t-1})_{j \in S_i^+}\by^\top_{t-1}   \left( \frac{1}{n} \sum_{t=2}^{n}  \by_{t-1} \epsilon_{i,t} \right)
		\end{array}
		\right)
	\end{eqnarray}	
	is asymptotically normal.
Let us consider one term in the upper block of (\ref{A.14}) first. For each $j\in S_i$, we have
\begin{align}\label{decom}
\frac{1}{n} \sum_{t=2}^{n}  y_{j,t}\by^\top_{t-1}  ( \frac{1}{n} \sum_{t=2}^{n}  \by_{t-1} \epsilon_{i,t})=&\frac{1}{n} \sum_{t=2}^{n}  (y_{j,t}\by^\top_{t-1}-E(y_{j,t}\by^\top_{t-1}))   \frac{1}{n} \sum_{t=2}^{n}  \by_{t-1} \epsilon_{i,t}\notag\\
&+\frac{n-1}{n}E(y_{j,t}\by^\top_{t-1})\frac{1}{n} \sum_{t=2}^{n}  \by_{t-1} \epsilon_{i,t}\notag\\
&=\frac{1}{n} \sum_{t=2}^{n}  (\be_j^\top\by_{t}\by^\top_{t-1}-\be_j^\top\bSigma_1)   \frac{1}{n} \sum_{t=2}^{n}  \by_{t-1} \epsilon_{i,t}\notag\\
&+\frac{n-1}{n}\be_j^\top\bSigma_1\frac{1}{n} \sum_{t=2}^{n}  \by_{t-1} \epsilon_{i,t}\notag\\
&=E_1+E_2.
\end{align}
By a similar argument as (14) in Dou et al. (2016), we can show that
\begin{equation}\label{rates}
E_1=O_p(\frac{p}{n})\quad\text{and}\quad E_2=O_p(\frac{1}{\sqrt{n}}).
\end{equation}
If $p=o(\sqrt{n})$, it follows that
\begin{equation}
\frac{1}{n} \sum_{t=2}^{n}  y_{j,t}\by^\top_{t-1}  ( \frac{1}{\sqrt{n}} \sum_{t=2}^{n}  \by_{t-1} \epsilon_{i,t})=\be_j^\top\bSigma_1\frac{1}{\sqrt{n}} \sum_{t=2}^{n}  \by_{t-1} \epsilon_{i,t}+o_p(1),\quad j\in S_i.
\end{equation}
Similarly, we can show that
\begin{equation}
\frac{1}{n} \sum_{t=2}^{n}  y_{j,t-1}\by^\top_{t-1}  ( \frac{1}{\sqrt{n}} \sum_{t=2}^{n}  \by_{t-1} \epsilon_{i,t})=\be_j^\top\bSigma_0\frac{1}{\sqrt{n}} \sum_{t=2}^{n}  \by_{t-1} \epsilon_{i,t}+o_p(1),\quad j\in S_i^+.
\end{equation}

Now it suffices to prove
$$S_{n,p} \equiv \vu_1^T\bI_{S_i}^\top\bSigma_1 \frac{1}{\sqrt{n}} \sum_{t=2}^{n}  \by_{t-1}\epsilon_{i,t} + \vu_2^\top \bI_{S_i^+}^\top\bSigma_0\frac{1}{\sqrt{n}} \sum_{t=2}^{n} \by_{t-1}\epsilon_{i,t} $$
is asymptotically normal,  where $\bI_{S_i}$ and $\bI_{S_i^+}$ are defined as those in (\ref{eqn:Vi}).
		
	Now we calculate the variance of $S_{n,p}$.	It holds that
\begin{align}\label{var1}
	\Var(\vu_1^T\bI_{S_i}^\top\bSigma_1 \frac{1}{\sqrt{n}} \sum_{t=2}^{n}  \by_{t-1}\epsilon_{i,t})=&\vu_1^T\bI_{S_i}^\top\bSigma_1\frac{n-1}{n}\bSigma_{\my,\bve_i}(0)\bSigma_1^\top\bI_{S_i}\bu_1\\
	&+\bu_1^\top\bI_{S_i}^\top\bSigma_1\sum_{j=1}^{n-2}(1-\frac{j+1}{n})[\bSigma_{\by,\bve_i}(j)+\bSigma^\top_{\by,\bve_i}(j)]\bSigma_1^\top\bI_{S_i}\bu_1.\notag
\end{align}
We note that
\begin{equation*}
E|\be_j^\top\bSigma_1\by_{t-1}\epsilon_{i,t}|^{\frac{4+\gamma}{2}}\leq[E|\be_j^\top\bSigma_1\by_{t-1}|^{4+\gamma}]^{\frac{1}{2}}[E|\epsilon_{i,t}|^{4+\gamma}]^{\frac{1}{2}}\leq \infty.
\end{equation*}
	By Proposition 2.5 of \cite{fan2003nonlinear}, it follows from $\sum_{j=1}^\infty\alpha_p(j)^{\frac{\gamma}{4+\gamma}}<\infty$ in Lemma \ref{lm1} that
\begin{align*}
\mathop{\sup}_{p}&\sum_{j=1}^{\infty}|\bu_1^\top\bI_{S_i}^\top\bSigma_1[\bSigma_{\by,\bve_i}(j)+\bSigma_{\by,\bve_i}^\top(j)]\bSigma_1^\top\bI_{S_i}\bu_1|\\
&\leq C\mathop{\sup}_{j_1,j_2\leq p}\sum_{j=1}^{\infty}|\be_{j_1}^\top\bSigma_1\bSigma_{\by,\bve_i}(j)\bSigma_1^\top\be_{j_2}|\\
&\leq C\mathop{\sup}_{l\leq p}\sum_{j=1}^{\infty}\alpha_p(j)^{\frac{\gamma}{4+\gamma}}(E|\be_{l}^\top\bSigma_1\by_{t-1}|^{4+\gamma})^{\frac{2}{4+\gamma}}(E|\epsilon_{i,t}|^{4+\gamma})^{\frac{2}{4+\gamma}} \;
< \; \infty.
\end{align*}
Similarly, 	
\begin{align*}
\Cov&\left(\vu_1^T\bI_{S_i}^\top\bSigma_1 \frac{1}{\sqrt{n}} \sum_{t=2}^{n}  \by_{t-1}\epsilon_{i,t},\vu_2^\top \bI_{S_i^+}^\top\bSigma_0\frac{1}{\sqrt{n}} \sum_{t=2}^{n} \by_{t-1}\epsilon_{i,t}\right)\\
=&\vu_1^T\bI_{S_i}^\top\bSigma_1\frac{n-1}{n}\bSigma_{\my,\bve_i}(0)\bSigma_0\bI_{S_i^+}\bu_2\notag\\
	&+\bu_1^\top\bI_{S_i}^\top\bSigma_1\sum_{j=1}^{n-2}(1-\frac{j+1}{n})[\bSigma_{\by,\bve_i}(j)+\bSigma^\top_{\by,\bve_i}(j)]\bSigma_0\bI_{S_i^+}\bu_2,
\end{align*}
and $\sup_p\sum_{j=1}^{\infty}|\bu_1^\top\bI_{S_i}^\top\bSigma_1[\bSigma_{\by,\bve_i}(j)+\bSigma_{\by,\bve_i}^\top(j)]\bSigma_0\bI_{S_i^+}\bu_2|<\infty$. Calculating all the variance and covariance and summing them up, it follows from dominate convergence theorem that
	\[ \textnormal{Var}\left( \frac{S_{n,p}}{\sqrt{\vu^\top \bU_i \vu}} \right) \rightarrow 1.  \]

		To prove the asymptotic normality of $S_{n,p}$, we can employ the small-block and large-block arguments  as those in Dou et al. (2016). We will borrow the notations $k_n$, $s_n$ and $l_n$ from their paper with the same properties and briefly introduce the steps for our case.
		%We partition the set $\{1, 2, ..., n\}$ into $2k_n + 1$ subsets with large blocks of size $l_n$, small blocks of size $s_n$ and the last remaining set of size $n - k_n(l_n + s_n)$. Put
%	\[ l_n = [ \sqrt{n} / \log n], \quad s_n = [\sqrt{n} \log n]^x,\quad k_n = [n/(l_n +s_n)], \]
%where $ \frac{\gamma}{4+\gamma} \leq x < 1$. Then,	
%	\[ l_n/\sqrt{n} \rightarrow  0, \quad s_n/l_n \rightarrow 0,\quad k_n = O(\sqrt{n} \log n). \]
%Note that $l_n/\sqrt{n} \rightarrow 0$ is important when we derive the Lindeberg condition of the truncated partial sum $T_{n,p}^L$ defined in \eqref{eqn:TnpL}.

%Since $  \sum_{l=1}^{\infty} \alpha_p(l)^{\frac{\gamma}{4+\gamma}} < \infty $, we have $\alpha_p (s_n) = o(s_n^{ -\frac{4+\gamma}{\gamma}}) $. Hence,
%	\begin{equation}
%%	\label{eqn:knalphasn}
%	  k_n \alpha_p (s_n) = o(k_n/ s_n^{ \frac{4+\gamma}{\gamma} } ) = o( \sqrt{n} \log n /  [ \sqrt{n} \log n]^{x \frac{4+\gamma}{\gamma} } ) \rightarrow 0.
	% \end{equation}
 We can partition $S_{n,p}$ in the following way
	\begin{eqnarray}
	S_{n,p} = && \vu_1^\top \frac{1}{\sqrt{n}} \sum_{j=1}^{k_n} \xi_j^{(1)} + \vu_2^\top \frac{1}{\sqrt{n}} \sum_{j=1}^{k_n} \xi_j^{(2)} + \vu_1^\top \frac{1}{\sqrt{n}} \sum_{j=1}^{k_n} \eta_j^{(1)} + \vu_2^\top \frac{1}{\sqrt{n}} \sum_{j=1}^{k_n} \eta_j^{(2)} \nonumber \\
	&& \vu_1^\top \frac{1}{\sqrt{n}} \zeta^{(1)} + \vu_2^\top  \frac{1}{\sqrt{n}} \zeta^{(2)},
	\end{eqnarray}
	where
	\[  \xi_j^{(1)} = \sum_{ t=(j-1)(l_n+s_n) + 1 }^{ jl_n + (j-1)s_n } \bI^\top_{S_i} \bSigma_1 \by_{t-1} \epsilon_{i,t}, \quad  \eta_j^{(1)} = \sum_{ t=jl_n +(j-1)s_n +1 }^{ j(l_n+s_n) } \bI^\top_{S_i} \bSigma_1 \my_{t-1} \epsilon_{i,t},  \]
	\[  \xi_j^{(2)} = \sum_{ t=(j-1)(l_n+s_n) + 1 }^{ jl_n + (j-1)s_n } \bI^\top_{S_i^+} \bSigma_0 \by_{t-1} \epsilon_{i,t}, \quad  \eta_j^{(2)} = \sum_{ t=jl_n +(j-1)s_n +1 }^{ j(l_n+s_n) } \bI^\top_{S_i^+} \bSigma_0 \my_{t-1} \epsilon_{i,t},  \]
	\[  \zeta^{(1)} = \sum_{ t=k_n(l_n+s_n) + 1 }^{ n } \bI^\top_{S_i} \bSigma_1 \my_{t-1} \epsilon_{i,t}, \quad  \zeta^{(2)} = \sum_{ t=k_n(l_n+s_n) +1 }^{ n } \bI^\top_{S_i^+} \bSigma_0 \by_{t-1} \epsilon_{i,t}, \]
	and the summation starts from $\by_0$ for the convenience of calculation.
	 Note that,  $\xi_j^{(1)}$, $\eta_j^{(1)}$ and $\zeta_j^{(1)}$ are $|S_i|$ dimensional vectors, and $\xi_j^{(2)}$, $\eta_j^{(2)}$ and $\zeta_j^{(2)}$  are $| S_i^+ |$ dimensional vectors.  Since $\alpha_p(n) = o( n^{ - \frac{(2+\gamma/2)2}{ 2(2+\gamma/2 -2) } } )$ and $k_n s_n/n \rightarrow 0$, $(l_n + s_n)/n \rightarrow 0$, by applying Proposition 2.7 of Fan and Yao (2003), it holds that
	\begin{equation}
	\label{eqn:smallterm}
	 \frac{1}{\sqrt{n}} \sum_{j=1}^{k_n} \eta_j^{(l)}=o_p(1)\quad \text{and}\quad  \frac{1}{\sqrt{n}} \zeta^{(l)}=o_p(1), \quad l=1,2.
	 \end{equation}
	Therefore,
	\begin{equation}
	\label{eqn:Tnp}
	S_{n,p} =  \vu_1^\top \frac{1}{\sqrt{n}} \sum_{j=1}^{k_n} \xi_j^{(1)} + \vu_2^\top \frac{1}{\sqrt{n}} \sum_{j=1}^{k_n} \xi_j^{(2)} + o_p(1)  \equiv T_{n,p} + o_p(1).
	\end{equation}
	 Similar to (\ref{var1}), we can calculate the variance of $ T_{n,p}$ and it holds that
	%\begin{eqnarray}
%	\label{eqn:VarofTnp1}
%	&& \Var \left( \vu_1^\top \frac{1}{\sqrt{n}} \sum_{j=1}^{k_n} \xi_j^{(1)}   \right) = \frac{k_n}{n} \vu_1^\top \Var \left( \xi_1^{(1)} \right) \vu_1 \{ 1+o(1) \} \nonumber \\
%	&& =  \frac{k_n}{n} \vu_1^\top \Var \left( \sum_{t=1}^{l_n} \bI^\top_{S_i} \bSigma_1 \by_{t-1} \epsilon_{i,t} \right) \vu_1 \{ 1+o(1) \} \nonumber \\
  %      && =  \frac{k_nl_n}{n} \vu_1^\top \Bigg\{ \bI^\top_{S_i} \bSigma_1 \bSigma_{\by, \bve_i}(0) \bSigma_1^\top \bI_{S_i} +   \sum_{j=1}^{l_n-1} (1-\frac{j}{l_n}) \bI^\top_{S_i} \bSigma_1\Big ( \bSigma_{\by,\bve_i}(j) + \bSigma^\top_{\by,\bve_i}(j)\Big )   \bSigma_1^\top \bI_{S_i}  \Bigg\}  \nonumber \\
     %   && \quad \times\vu_1 \{ 1+o(1) \} .
	%\end{eqnarray}
%	Calculating all the variance and covariance and summing up them, it follows from dominate convergence theorem and $\frac{k_n l_n}{n} \rightarrow 1$ that
\begin{equation}\label{varoft}
	 \Var \left(  \frac{T_{n,p}}{ \sqrt{\vu^\top \bU_i \vu }} \right) \rightarrow 1,
	\end{equation}
	see also Dou et al. (2016) for a similar argument. Now, it suffices to prove the asymptotic normality of $T_{n,p}$. We partition $T_{n,p}$ into two parts via truncation. Specifically, we define
		\[   \xi_j^{(1)L} =  \sum_{ t=(j-1)(l_n+s_n) + 1 }^{ jl_n + (j-1)s_n } \bI^\top_{S_i} \bSigma_1 \my_{t-1} \epsilon_{i,t} \bI_{ \{ \|\bI^\top_{S_i} \bSigma_1\by_{t-1} \epsilon_{i,t} \|_2 \leq L \} }  , \]
		and
	\[   \xi_j^{(1)R} =  \sum_{ t=(j-1)(l_n+s_n) + 1 }^{ jl_n + (j-1)s_n } \bI^\top_{S_i} \bSigma_1 \by_{t-1} \epsilon_{i,t} \bI_{ \{\| \bI^\top_{S_i} \bSigma_1\by_{t-1} \epsilon_{i,t} \|_2 > L \} }  . \]
	 Similarly, we can define $ \xi_j^{(2)L} $ and $ \xi_j^{(2)R} $. Then,
	 \begin{eqnarray}
	 \label{eqn:TnpL}
	 T_{n,p} & =  & \left(  \vu_1^T \frac{1}{\sqrt{n}} \sum_{j=1}^{k_n} \xi_j^{(1)L} + \vu_2^T \frac{1}{\sqrt{n}} \sum_{j=1}^{k_n} \xi_j^{(2)L}  \right)  + \left(  \vu_1^T \frac{1}{\sqrt{n}} \sum_{j=1}^{k_n} \xi_j^{(1)R} + \vu_2^T \frac{1}{\sqrt{n}} \sum_{j=1}^{k_n} \xi_j^{(2)R}  \right)  \nonumber \\
	 & \equiv &  T_{n,p}^L + T_{n,p}^R .
 	 \end{eqnarray}
	 Define
$$   \bSigma_{\by,\bve_i,L}^{(S_i,S_i^+)}(j) = \textnormal{Cov}(\my_{t-1+j}\epsilon_{i,t+j} \bI_{ \{ \| \bI^\top_{S_i} \bSigma_1 \by_{t-1+j} \epsilon_{i,t+j} \|_2 \leq L \} } , \by_{t-1}\epsilon_{i,t} \bI_{ \{ \|\bI^\top_{S_i^+} \bSigma_0  \my_{t-1} \epsilon_{i,t} \|_2 \leq L \} } )$$
for $j=0,1,2,...$, and
	 \[   \bSigma^{(S_i,S_i^+)}_{\by,\bve_i,L} = \bSigma^{(S_i,S_i^+)}_{\my,\bve_i,L}(0) + \sum_{j=1}^\infty \left( \bSigma^{(S_i,S_i^+)}_{\by,\bve_i,L}(j)  + (\bSigma^{(S_i,S_i^+)}_{\by,\bve_i,L}(j))^\top \right). \]
	 Similarly we have $\bSigma^{(S_i,S_i)}_{\by,\bve_i,L}$, $\bSigma^{(S_i^+,S_i)}_{\by,\bve_i,L}$ and $\bSigma^{(S_i^+,S_i^+)}_{\by,\bve_i,L}$. Let
	 \begin{eqnarray}
	  \bU_i^L \equiv \left( \begin{array}{ll} \bI_{S_i}^\top \bSigma_1 \bSigma^{(S_i,S_i)}_{\by,\bve_i,L} \bSigma_1^\top \bI_{S_i} & \bI_{S_i}^\top \bSigma_1 \bSigma^{(S_i,S_i^+)}_{\by,\bve_i,L} \bSigma_0 \bI_{S_i^+} . \\ \bI_{S_i^+}^\top \bSigma_0 \bSigma^{(S_i^+,S_i)}_{\by,\bve_i,L} \bSigma_1^\top \bI_{S_i} & \bI_{S_i^+}^\top \bSigma_0 \bSigma^{(S_i^+,S_i^+)}_{\by,\bve_i,L} \bSigma_0 \bI_{S_i^+}
	  \end{array}  \right).
	  \end{eqnarray}
	 Then $ \bU_i^L  \rightarrow \bU_i $ as $L \rightarrow \infty$.
	Similar to  \eqref{varoft}, it holds that
	\[  \Var \left(  \frac{T_{n,p}^L}{\sqrt{\vu^\top \bU_i^L \vu}}  \right) \rightarrow 1.   \]
	If we define $U_i^R$ in a similar way, then $U_i^R \rightarrow 0$ as $L\rightarrow\infty$ and
	 $ \Var ( {T_{n,p}^R}/{\sqrt{\vu^\top \bU_i^R \vu} } ) \rightarrow 1$ as $n\rightarrow\infty$.
	Define
	\begin{equation}
	M_{n,p} = \left |  E \exp \left(  \frac{itT_{n,p}}{\sqrt{\vu^\top\bU_i\vu}} \right)  - \exp \left(  - \frac{t^2}{2}  \right)   \right |,
	\end{equation}
	where $i=\sqrt{-1}$. Then, the required result follows from the statement that
	\begin{equation}
	\label{eqn:Mnplimit}
	   \lim_{n \rightarrow \infty} M_{n,p} < \delta,
	\end{equation}
	for any given $\delta >0$. This can be done by following the same arguments as part 2.7.7 of Fan and Yao (2003), see also Dou et al. (2016).
	Therefore, the proof of assertion (1) is completed.

	To prove assertion (2),  it is sufficient to show that each element of $\wh \bV_i^\top \wh\bV_i$ converges in probability to the corresponding element of $\bK_i$. By (\ref{b6}), we have
	
	 \begin{eqnarray*}
	  \wh \bV_i^\top\wh\bV_i \equiv \left( \begin{array}{ll} \frac{1}{n}\sum\limits_{t=2}^n\bI_{S_i}^\top \by_t\by_{t-1}^\top \frac{1}{n}\sum\limits_{t=2}^{n}\by_{t-1}\by_{t}^\top\bI_{S_i} & \frac{1}{n}\sum\limits_{t=2}^n\bI_{S_i}^\top \by_t\by_{t-1}^\top \frac{1}{n}\sum\limits_{t=2}^{n}\by_{t-1}\by_{t-1}^\top\bI_{S_i^+}\\ \frac{1}{n}\sum\limits_{t=2}^n\bI_{S_i^+}^\top \by_{t-1}\by_{t-1}^\top \frac{1}{n}\sum\limits_{t=2}^{n}\by_{t-1}\by_{t}^\top\bI_{S_i} & \frac{1}{n}\sum\limits_{t=2}^n\bI_{S_i^+}^\top \by_{t-1}\by_{t-1}^\top \frac{1}{n}\sum\limits_{t=2}^{n}\by_{t-1}\by_{t-1}^\top\bI_{S_i^+}
	  \end{array}  \right).
	  \end{eqnarray*}
Let us take one element of $ \wh \bV_i^\top\wh\bV_i $ as an example. For some $j_1,j_2\in S_i$,
\begin{align}\label{vv}
\frac{1}{n} &\sum_{t=1}^{n}\be_{j_1}^\top\by_t \by^\top_{t-1} \frac{1}{n} \sum_{t=1}^{n} \by_{t-1} \by_{t}^\top\be_{j_2}\notag\\
=&\left(\frac{1}{n} \sum_{t=1}^{n}\be_{j_1}^\top\by_t \by^\top_{t-1}-\be_{j_1}^\top\bSigma_1\right)\left(\frac{1}{n} \sum_{t=1}^{n} \by_{t-1} \by_{t}^\top\be_{j_2}-\bSigma_1^\top\be_{j_2}\right)\notag\\
&+\be_{j_1}^\top\bSigma_1\left(\frac{1}{n} \sum_{t=1}^{n} \by_{t-1} \by_{t}^\top\be_{j_2}-\bSigma_1^\top\be_{j_2}\right)
+\left(\frac{1}{n} \sum_{t=1}^{n}\be_{j_1}^\top\by_t \by^\top_{t-1}-\be_{j_1}^\top\bSigma_1\right)\bSigma_1^\top\be_{j_2}\notag\\
&+\be_{j_1}^\top\bSigma_1\bSigma_1^\top\be_{j_2}.
\end{align}	
	%\begin{eqnarray}
%	&& \frac{1}{n} \sum_{t=1}^{n} \by^\top_{t-1} y_{j,t} \frac{1}{n} \sum_{t=1}^{n} \by_{t-1} y_{j,t} \nonumber \\
%	= && \left( \frac{1}{n} \sum_{t=1}^{n} \by^\top_{t-1} y_{j,t} - \be_j^\top \bSigma_1 \right) \left( \frac{1}{n} \sum_{t=1}^{n} \by_{t-1} y_{j,t} -\bSigma_1^\top \be_j \right)  \nonumber \\
%	&& + 2 \be_j^\top\bSigma_1  \left( \frac{1}{n} \sum_{t=1}^{n} \by_{t-1} y_{j,t} -\bSigma_1^\top \be_j \right)  + \be_j^\top \bSigma_1 \bSigma_1^\top \be_j.
%	\end{eqnarray}
Using the same arguments as (\ref{rates}), the first term is $O_p(\frac{p}{n})$ and the second and the third terms are of order $O_p(\frac{1}{\sqrt{n}})$. Hence given $p=o(n)$, it holds that
	\[  \frac{1}{n} \sum_{t=1}^{n}\be_{j_1}^\top\by_t \by^\top_{t-1} \frac{1}{n} \sum_{t=1}^{n} \by_{t-1} \by_{t}^\top\be_{j_2} / (\be_{j_1}^\top \bSigma_1 \bSigma_1^\top \be_{j_2}) \rightarrow 1.  \]
Applying the same arguments to the other elements of $\wh\bV_i^\top \wh\bV_i$, we have
	\[ \bK_i (\wh\bV_i^\top \wh\bV_i)^{-1} \rightarrow_p \bI_{\tau_i} .\]

{When $k_0$ is diverging with the rate $o(C_n^{-1}n/\log(p\vee n))$, Theorem 1 still holds. We can also show that $\|\wh\bV_i^\top\wh\bV_i-\bK_i\|_F=O_p(\sqrt{\frac{k_0^2}{n}})=o_p(1)$ if $p=o(\sqrt{n})$ since $k_0<p$, then we have $\lambda_{\min}(\wh\bV_i^\top \wh\bV_i)\geq c$ with probability tending to 1. By (\ref{b5}), (\ref{decom}) and (\ref{rates}),}

$$\|\wh \bbeta_i - \bbeta_i\|_2 \leq C\|{1 \over n}\wh \bV_i^\top \sum_{t=2}^n \by_{t-1} \epsilon_{i,t}\|_2=O_p(\sqrt{\frac{k_0}{n}}),
\quad i =1,..., p.$$
	
	Part $(ii)$ of Theorem 2 for a fixed $k_0$ follows immediately from (\ref{decom}) and (\ref{vv}) if $\sqrt{n}=O(p)$ and $p=o(n)$.
	
{When $k_0$ is diverging with the rate $o\{\min(C_n^{-1}n/\log(p\vee n),n/p)\}$, by a similar argument as above, we have $\|\wh\bV_i^\top\wh\bV_i-\bK_i\|_F=O_p(\sqrt{\frac{k_0^2p^2}{n^2}})=o_p(1)$, and $\lambda_{\min}(\wh\bV_i^\top \wh\bV_i)\geq c$ with probability tending to 1.  If $\sqrt{n}=O(p)$ and $p=o(n)$, by (\ref{b5}), (\ref{decom}) and (\ref{rates}), }
	$$\|\wh \bbeta_i - \bbeta_i\|_2 \leq C\|{1 \over n}\wh \bV_i^\top \sum_{t=2}^n \by_{t-1} \epsilon_{i,t}\|_2=O_p(\sqrt{\frac{k_0p^2}{n^2}})=o_p(k_0^{1/2}p/n),
\quad i =1,..., p.$$
The proof is completed. $\Box$

\begin{landscape}

\bigskip
\bigskip

\begin{table}[!h]
\renewcommand\arraystretch{1.5}
\bc
\caption{\label{tab:t1} Relative frequencies (\%) of the occurrence of
the events $ \{\hat k=k_0\}$, $ \{\hat k>k_0\}$  and $ \{\hat k<k_0\}$ based on $r=1$,
and mean and  standard deviations (in parentheses) of $ \|\bA-\hat \bA\|_2 $
and $ \|\bB-\hat \bB\|_2 $ for Case 1 with $k_0=3$, $K=10$ and $r=1,2, 3$ respectively.}
\vspace{0.25cm}
{\small
\begin{tabular}{ccc| ccc|cc|cc|cc}
\hline
   &     &     &     \multicolumn{3}{|c}{$r=1$ } & \multicolumn{2}{|c}{$r=1$}& \multicolumn{2}{|c}{$r=2$}& \multicolumn{2}{|c}{$r=3$} \\
$p$& $n$ & SNR &  $ \{\hat k=k_0\} $ & $\{\hat k>k_0\} $ & $\{\hat k<k_0\} $&$ \|\bA-\hat \bA\|_2 $ &$ \|\bB-\hat \bB\|_2 $&$ \|\bA-\hat \bA\|_2 $ &$ \|\bB-\hat \bB\|_2 $ &$ \|\bA-\hat \bA\|_2 $ &$ \|\bB-\hat \bB\|_2 $ \\
\hline
 100  &     500 &       1.136 &       0.460 &       0.540 &       0.000 &       1.124  (0.414)&       0.525 (0.191) &   1.042 (0.334) & 0.613 (0.138) &  1.028 (0.310) & 0.799 (0.135)   \\
      &    1,000&       1.136 &       0.972 &       0.028 &       0.000 &       0.670 (0.116) &       0.269 (0.043) &   0.677 (0.102) & 0.345 (0.041) &  0.693 (0.096) & 0.470 (0.043)  \\
      &    2,000 &       1.136 &       1.000 &       0.000 &       0.000 &       0.576 (0.062) &       0.204 (0.018)&    0.590 (0.053) & 0.243 (0.021)&  0.613 (0.050) & 0.315 (0.025) \\
\hline
300  &     500  &       1.061 &       0.006 &       0.994 &       0.000 &       1.132 (0.116) &  0.619 (0.072) &  1.141 (0.116)  & 1.004 (0.077) &  1.163 (0.117)  & 1.337 (0.089) \\
     &    1,000 &       1.061 &       0.528 &       0.472 &       0.000 &       0.788 (0.168) &  0.319 (0.076) &  0.799 (0.163)  & 0.620 (0.082) &  0.818 (0.161)  & 0.896 (0.101)\\
     &    2,000 &       1.061 &       0.972 &       0.028 &       0.000 &       0.614 (0.050) &  0.198 (0.017) &   0.635 (0.048) & 0.387 (0.024) &  0.652 (0.048)  & 0.593 (0.029) \\
 \hline
500 &     500   &       1.112 &       0.034 &       0.966 &       0.000 &       1.082 (0.124)  &      0.617 (0.085)  & 1.107 (0.126) & 1.138 (0.098) & 1.142 (0.130) & 1.509 (0.112)\\
    &   1,000   &       1.112 &       0.552 &       0.448 &       0.000 &       0.812 (0.138) &       0.360 (0.068)  & 0.829 (0.142) & 0.791 (0.102) & 0.860 (0.148) & 1.129 (0.134) \\
    &   2,000  &       1.112 &       0.966 &       0.034 &       0.000 &       0.674 (0.058)  &      0.252 (0.024)   & 0.695 (0.058) & 0.543 (0.039) & 0.721 (0.058) & 0.816 (0.053) \\
 \hline
 800 &   500   &       1.166 &       0.368 &       0.632 &       0.000 &       0.820 (0.149) &      0.495 (0.098) & 0.843 (0.162) & 1.016 (0.127) & 0.879 (0.171) & 1.347 (0.150)   \\
     &   1,000 &       1.166 &       0.942 &       0.058 &       0.000 &       0.619 (0.058) &     0.319 (0.033)  & 0.640 (0.061) & 0.759 (0.049) & 0.671 (0.067) & 1.065 (0.061) \\
     &   2,000 &       1.166 &       0.998 &       0.002 &       0.000 &       0.569 (0.022) &     0.261 (0.014)  & 0.594 (0.024) & 0.598 (0.020) & 0.625 (0.025) & 0.878 (0.023) \\
\hline
1,000 &     500 &       1.054 &       0.000 &       1.000 &       0.000 &       0.851 (0.045) &     0.591 (0.025) &0.896 (0.045)  & 1.076 (0.044) & 0.932 (0.048) & 1.382 (0.049)\\
      &    1,000 &       1.054 &       0.242 &       0.758 &       0.000 &       0.620 (0.090) &    0.335 (0.062) & 0.653 (0.096) & 0.799 (0.067) & 0.686 (0.098) & 1.100 (0.077)\\
      &    2,000 &       1.054 &       0.984 &       0.016 &       0.000 &       0.472 (0.023) &    0.187 (0.013) & 0.495 (0.023) & 0.562 (0.018) & 0.522 (0.025) & 0.837 (0.021)\\
\hline
\end{tabular}}
\ec
\end{table}

\bigskip
\bigskip

\begin{table}[!h]
\renewcommand\arraystretch{1.5}
\bc
\caption{\label{tab:t2} Relative frequencies (\%) of the occurrence of
the events $ \{\hat k=k_0\}$, $ \{\hat k>k_0\}$  and $ \{\hat k<k_0\}$ based on $r=1$,
and mean and  standard deviations (in parentheses) of $ \|\bA-\hat \bA\|_2 $
and $ \|\bB-\hat \bB\|_2 $ for Case 2 with $k_0=3$, $K=10$ and $r=1,2, 3$ respectively.}
\vspace{0.25cm}
{\small
\begin{tabular}{ccc| ccc|cc|cc|cc}
\hline
   &     &     &     \multicolumn{3}{|c}{$r=1$ } & \multicolumn{2}{|c}{$r=1$}& \multicolumn{2}{|c}{$r=2$}& \multicolumn{2}{|c}{$r=3$} \\
$p$& $n$ & SNR &  $ \{\hat k=k_0\} $ & $\{\hat k>k_0\} $ & $\{\hat k<k_0\} $&$ \|\bA-\hat \bA\|_2 $ &$ \|\bB-\hat \bB\|_2 $&$ \|\bA-\hat \bA\|_2 $ &$ \|\bB-\hat \bB\|_2 $ &$ \|\bA-\hat \bA\|_2 $ &$ \|\bB-\hat \bB\|_2 $ \\
\hline
  100  &     500  &       1.068 &       0.014 &       0.986 &       0.000 &       1.672 (0.242) &      0.700 (0.096) & 1.465 (0.189) &   0.704 (0.067)  &  1.408 (0.176) &       0.878 (0.070)   \\
       &    1,000 &       1.068 &       0.520 &       0.480 &       0.000 &       1.028 (0.412) &      0.347 (0.135) & 0.960 (0.336) &   0.387 (0.086)  & 0.948 (0.313)  &       0.497 (0.074) \\
       &    2,000 &       1.068 &       0.976 &       0.024 &       0.000 &       0.628 (0.098) &      0.182 (0.027) & 0.638 (0.083) &   0.220 (0.023)  & 0.655 (0.078)  &       0.287 (0.023)  \\
\hline
   300 &     500  &       1.094 &       0.188 &       0.812 &       0.000 &       0.860 (0.185) &   0.504 (0.115) & 0.870 (0.181)  &   0.820 (0.118) &  0.894 (0.180) & 1.116 (0.139)   \\
       &    1,000 &       1.094 &       0.896 &       0.104 &       0.000 &       0.561 (0.100) &   0.258 (0.044)  &0.570 (0.092) &  0.492 (0.046) & 0.590 (0.089) &  0.727 (0.056) \\
       &    2,000 &       1.094 &       0.990 &       0.010 &       0.000 &       0.484 (0.034) &   0.183 (0.015)  &0.504 (0.035) &  0.328 (0.019) & 0.523 (0.035) &  0.504 (0.023)  \\
\hline
   500 &     500  &       1.215 &       0.762 &       0.238 &       0.000 &       0.689 (0.125) &       0.428 (0.083)  &  0.700 (0.129)    &   0.829 (0.096) & 0.729 (0.133) & 1.121 (0.112) \\
       &    1,000 &       1.215 &       0.988 &       0.012 &       0.000 &       0.572 (0.037) &       0.309 (0.024)  &  0.590 (0.038)    &   0.637 (0.035) & 0.620 (0.039) & 0.908 (0.041) \\
       &    2,000 &       1.215 &       1.000 &       0.000 &       0.000 &       0.516 (0.025 )&       0.249 (0.014)  &  0.543 (0.024)    &   0.477 (0.021) & 0.574 (0.026) & 0.710 (0.026) \\
\hline
   800 &     500 &        1.258 &       0.998 &       0.002 &       0.000 &       0.491 (0.025) &       0.349 (0.021) &   0.500 (0.025)  & 0.704 (0.027)  & 0.543 (0.026) & 0.968 (0.030)  \\
       &    1,000 &       1.258 &       1.000 &       0.000 &       0.000 &       0.432 (0.017) &       0.268 (0.015) &   0.447 (0.018)  & 0.582 (0.021)  & 0.493 (0.021) & 0.842 (0.024)  \\
       &    2,000 &       1.258 &       1.000 &       0.000 &       0.000 &       0.386 (0.015) &       0.212 (0.011) &    0.408 (0.016) & 0.450 (0.016)  & 0.448 (0.018) & 0.683 (0.020) \\
\hline
  1,000 &     500 &        1.064 &       0.000 &       1.000 &       0.000 &       0.948 (0.052)  &       0.610 (0.033) & 0.997 (0.055) & 1.160 (0.049) & 1.031 (0.056) & 1.498 (0.054)   \\
        &    1,000 &       1.064 &       0.218 &       0.782 &       0.000 &       0.720 (0.095)  &       0.359 (0.061) & 0.752 (0.104) & 0.886 (0.083) & 0.786 (0.110) & 1.217 (0.102)   \\
        &    2,000 &       1.064 &       0.916 &       0.084 &       0.000 &       0.556 (0.044)  &        0.213 (0.023) &0.577 (0.049) & 0.621 (0.040) & 0.602 (0.054) & 0.913 (0.053)  \\
\hline
\end{tabular}}
\ec
\end{table}

\begin{table}[!h]
\renewcommand\arraystretch{1.5}
\bc
\caption{\label{tab:t6} Relative frequencies (\%) of the occurrence of
the events $ \{\hat k=k_0\}$, $ \{\hat k>k_0\}$  and $ \{\hat k<k_0\}$ based on   $r=1$,
and mean and  standard deviations (in parentheses) of $ \|\bA-\hat \bA\|_2 $
and $ \|\bB-\hat \bB\|_2 $ for Case 2 with $k_0=1$, $d_i=\min(p,\lbrack n^{0.495}\rbrack)$ and  $K=5$.}
\vspace{0.25cm}
{\small
\begin{tabular}{ccc| ccc|cc|cc}
\hline
   &     &     &     \multicolumn{3}{|c}{$r=1$ } & \multicolumn{2}{|c}{Estimator I}& \multicolumn{2}{|c}{ Estimator II}\\\
$p$& $n$ & SNR &  $ \{\hat k=k_0\} $ & $\{\hat k>k_0\} $ & $\{\hat k<k_0\} $&$ \|\bA-\hat \bA\|_2 $ &$ \|\bB-\hat \bB\|_2 $&$ \|\bA-\hat \bA\|_2 $ &$ \|\bB-\hat \bB\|_2 $ \\
\hline
50 & 2,500  &   1.379 &  0.956 &   0.044 &  0.000 &  0.543 (0.244) &  0.201 (0.086) &  0.570 (0.254) &  0.204 (0.096)  \\
   & 5,000  &   1.378 &  0.998 &   0.002 &  0.000 &  0.417 (0.130) &  0.132 (0.026) &  0.417 (0.130) &  0.132 (0.026)  \\
   & 1,0000 &  1.379  &  1.000 &   0.000 &  0.000 &  0.396 (0.154) &  0.100 (0.024) &  0.396 (0.154) &  0.100 (0.024)  \\
\hline
75 & 2,500  &  1.321 &  1.000   &  0.000 &  0.000 &  0.358 (0.088) &   0.170 (0.034) &  0.597 (0.086) &  0.426 (0.099)  \\
   & 5,000  &  1.320 &  1.000   &  0.000 &  0.000 &  0.326 (0.106) &   0.143 (0.041) &  0.415 (0.087) &  0.166 (0.037)  \\
   & 1,0000 &  1.320 &  1.000   &  0.000 &  0.000 &  0.313 (0.117) &   0.125 (0.045) &  0.313 (0.117) &  0.125 (0.045)   \\
\hline
100 & 2,500  &   1.405 &  0.994  &  0.006 &  0.000 &  0.417 (0.076) &   0.215 (0.035) &  0.765 (0.106) &  0.663 (0.121)   \\
    & 5,000  &   1.405 &  0.998  &  0.002 &  0.000 &  0.345 (0.075) &   0.160 (0.029) &  0.639 (0.092) &  0.514 (0.098)  \\
    & 1,0000 &   1.405 &  1.000  &  0.000 &  0.000 &  0.300 (0.090) &   0.122 (0.027) &  0.394 (0.087) &  0.156 (0.046)  \\
\hline
125 & 2,500  &    1.446 &  0.998 &   0.002 &  0.000 &  0.429 (0.065) &   0.215 (0.032) &  0.828 (0.100) &  0.764 (0.113)  \\
    & 5,000  &    1.446 &  1.000 &   0.000 &  0.000 &  0.380 (0.088) &   0.157 (0.023) &  0.690 (0.097) &  0.588 (0.093)\\
    & 1,0000 &    1.446 &  1.000 &   0.000 &  0.000 &  0.356 (0.100) &   0.112 (0.018) &  0.532 (0.083) &  0.231 (0.068) \\
\hline
\end{tabular}}
\ec
\end{table}

\end{landscape}

\begin{table}[!h]
 \renewcommand\arraystretch{1.5}
\bc
\caption{\label{tab:t3} Example 1 -- one-step and two-step ahead
post-sample mean squared predictive errors and their standard
deviations (in parentheses) over the 36 stations.}
\vspace{0.25cm}
{\small
\begin{tabular}{l|c| c|c}
Ordering & $\hat k$ &   One-step ahead &Two-step ahead \\
\hline
north to south & 5&             0.108 (0.283) &  0.161 (0.455)\\
west to east   & 5 &              0.107 (0.280) & 0.161 (0.309) \\
northwest to southeast & 7 &    0.223 (0.483) & 0.325 (0.690)\\
northeast to southwest  & 7 &   0.154 (0.435)  & 0.215 (0.452)\\
distance to Miyun & 5&   0.107 (0.315) &   0.190 (0.577) \\
\end{tabular}} \label{t3}
\ec
\end{table}

\begin{table}[!h]
\renewcommand\arraystretch{1.5}
\bc
\caption{\label{tab:t4} Example 1 -- percentages of correct one-step
ahead and two-step ahead predictions at the 7 different pollution levels across 36 stations.}
\vspace{0.25cm}
{\small
\begin{tabular}{lr|ccccccc}
Ordering&& Level 1  &  Level 2  &  Level 3  & Level 4  & Level 5  & Level 6  & Level 7   \\
\hline
north to south   & 1-step &      71.8 &   69.7 &     70.8 &    73.8 &   84.5 &   100 &   100 \\
                  & 2-step &     68.9 &   66.4 &     68.7 &    73.4 &   84.1 &  100 &    100 \\
west to east      & 1-step &     76.2 &   69.7 &     66.8 &    77.3 &   87.8 &  100  &   100 \\
                  & 2-step &     72.1 &   64.4 &     62.1 &    75.3 &   86.3 &  100 &   100 \\
 NW to SE         & 1-step &     72.4 &   66.5 &     61.3 &    71.3 &   87.1 &  100 &   100  \\
                  & 2-step &     68.9 &   63.2 &     59.0 &    68.5 &   86.1 &  100 &   100   \\
 NE to SW         &1-step &      75.1 &   62.4 &     63.6 &    73.5 &   87.1 &  100 &   100\\
                  &2-step &      71.1 &  59.8  &     60.4 &    72.7 &   86.7 &  100 &   100 \\
distance to Miyun &1-step &      73.4 &   72.8 &     67.9 &    72.7 &   85.9 &   100&   100 \\
                  &2-step &      68.6 &   67.7 &     62.6 &    71.2 &   85.7 &   100&   100 \\
\end{tabular}}\label{t4}
\ec
\end{table}

\begin{table}[!h]
 \renewcommand\arraystretch{1.5}
\bc
\caption{\label{tab:t5} Example 2 -- one-step
and two-step ahead  post-sample mean squared predictive errors  over 41 components and their standard
deviations (in parentheses).}
\vspace{0.25cm}
{\small
\begin{tabular}{l| c|c}
  &   One-step ahead &Two-step ahead \\
\hline
Banded Model with $\wh k=1$  &     0.001 (0.001)  & 0.020 (0.056)\\
Dou et al's model with distance weights   &   0.001 (0.001) &  3.229 ( 6.468)\\
Dou et al's  model with correlation weights  &     0.008 (0.020) &   1.107 (0.930)\\
\end{tabular}} \label{t5}
\ec
\end{table}

\begin{figure}[h]
\begin{center}
\includegraphics[width=6 in]{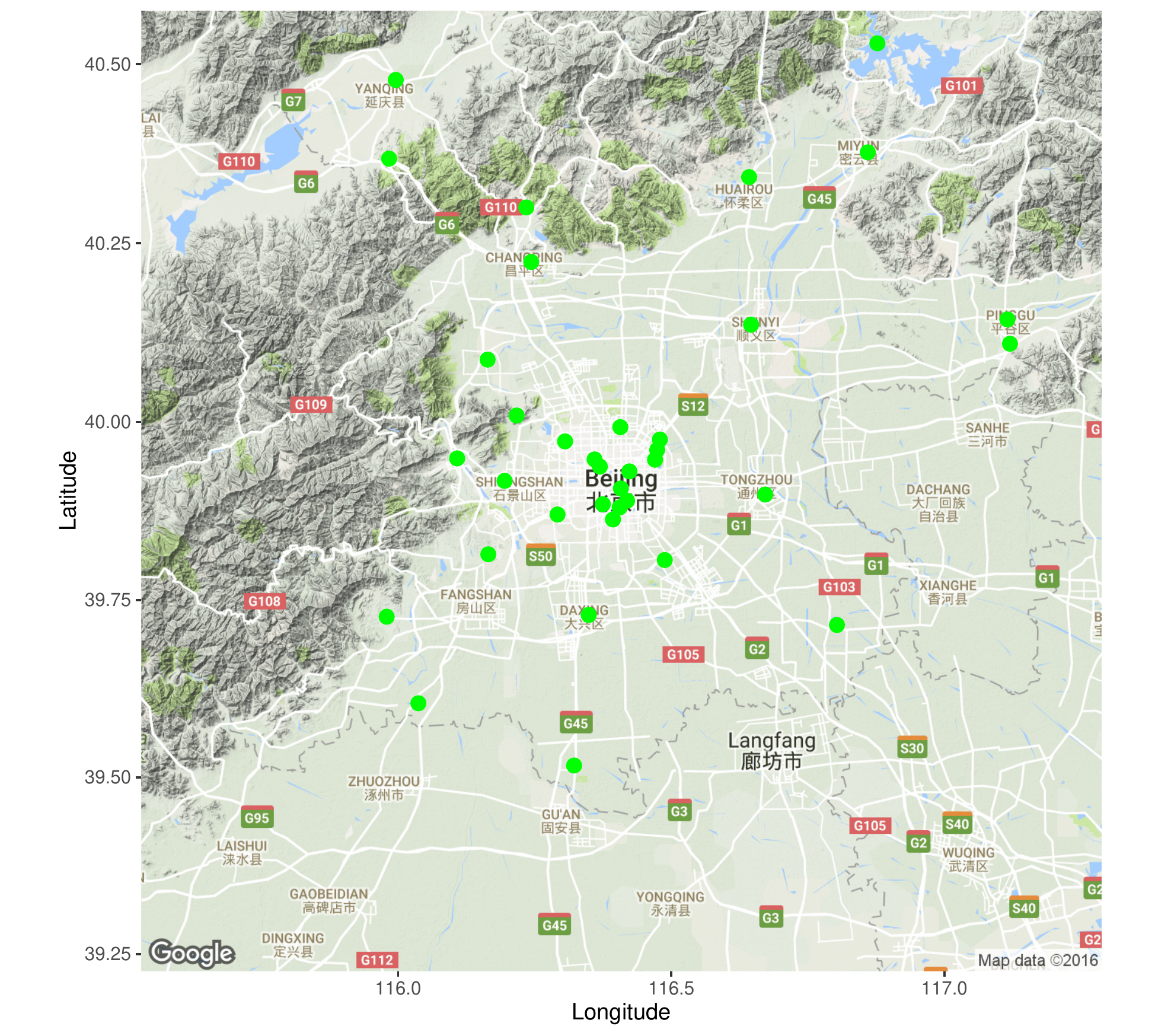}
\caption{Map of the 36 PM$_{2.5}$ monitoring stations in Beijing} \label{fig1}
\end{center}
\end{figure}

\begin{figure}[h]
\begin{center}
\includegraphics[height=3.5in]{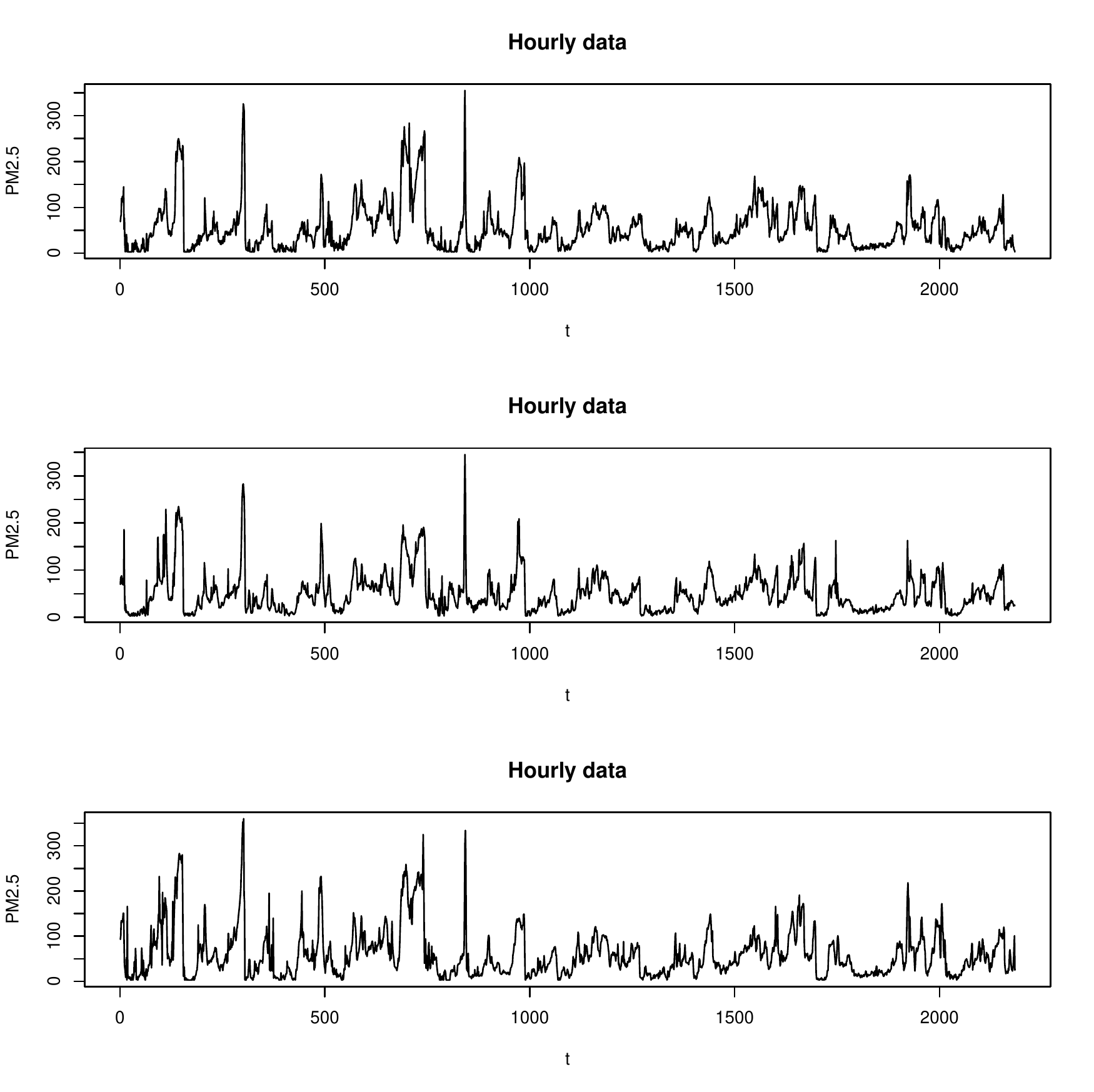}
\caption{Time series plots of hourly  $\mbox{PM}_{2.5}$  readings  in the
period of 1 April -- 30 June 2016 at,  from top to bottom,  MiYun, Huairou and Shunyi.}
\label{fig2}
\includegraphics[height=3.5in]{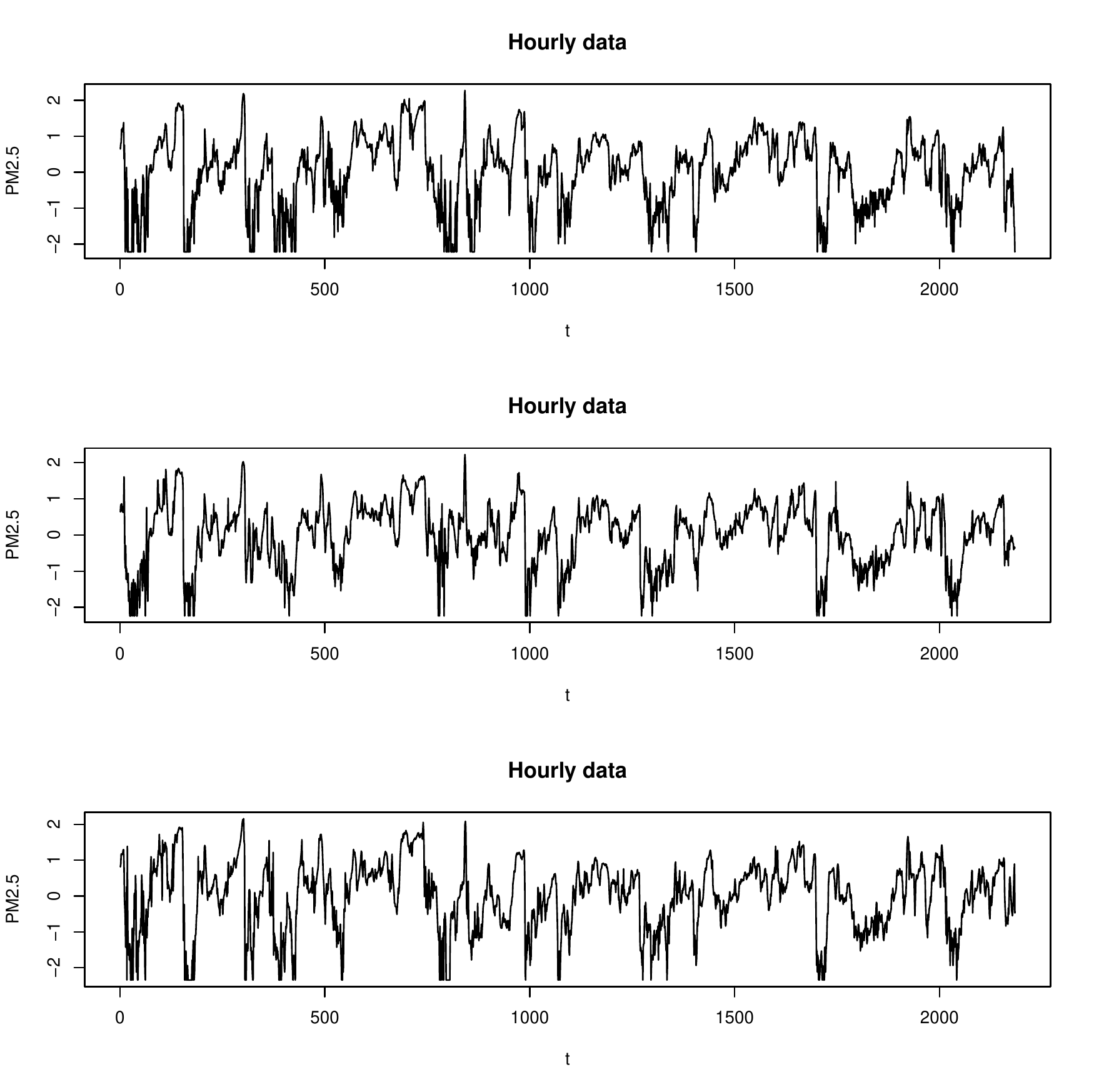}
\caption{Time series plots of the log-transformed and  centered hourly  $\mbox{PM}_{2.5}$ readings
in the period of 1 April -- 30 June 2016 at,  from top to bottom,
 MiYun, Huairou and Shunyi.} \label{fig3}
\end{center}
\end{figure}

\begin{figure}[h]
\begin{center}
\includegraphics[ height=3.5in]{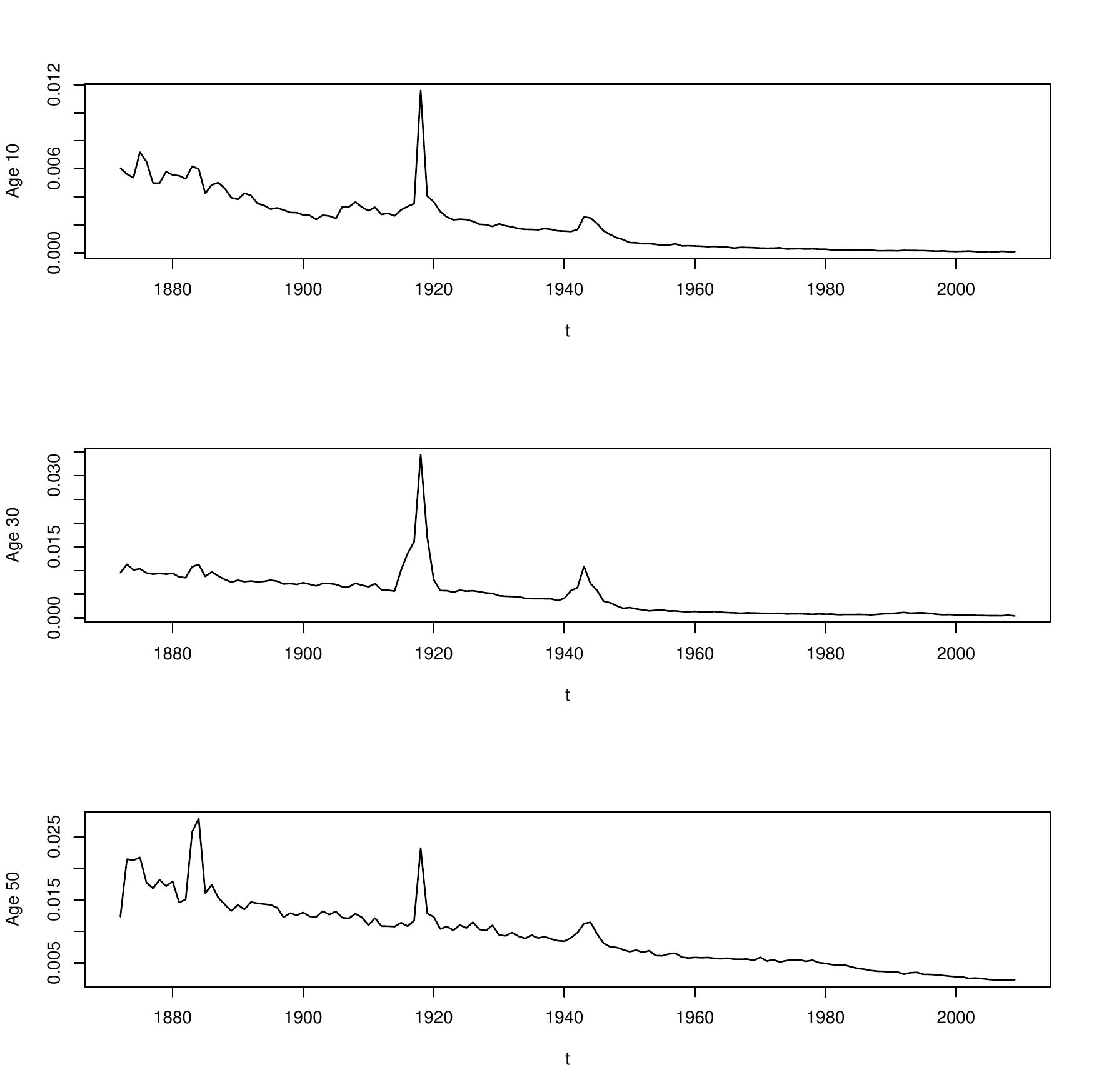}
\caption{Time series plots of the original yearly mortality rates (male and female in total) in the
period of 1951 -- 2009 for ages $i=10,30,50$.}
\label{fig4}
%\includegraphics[height=3.5in]{fig5.pdf}
%\caption{Time series plots of the log-scaled yearly mortality rates (male and female in total) in percentage after centering from both $n$ and $p$ dimensions    in the
%period of 1951 -- 2009 for ages $i=10,30,50$.} \label{fig5}
\end{center}
\end{figure}
%\bibliography{bsta.bib}
%\bibliographystyle{dcu}

\end{document}